\documentclass[preprint, twocolumn]{aastex7}
\usepackage{graphicx}
\usepackage{booktabs}

\usepackage{hyperref}

\begin{document}

\title{\texttt{XSNAP}: An X-ray Supernova Analysis Pipeline with Application to the Type II Supernova 2024ggi}
\author[orcid=0009-0008-7581-3096, sname='', gname='Ferdinand']{Ferdinand}
\affiliation{Department of Astronomy, University of Illinois at Urbana-Champaign, Urbana, IL 61801, USA}
\email[show]{ff10@illinois.edu}

\author[orcid=0000-0002-3934-2644, sname=Jacobson-Galán,gname=Wynn]{W. V. Jacobson-Galán}
\altaffiliation{NASA Hubble Fellow}
\affiliation{Cahill Center for Astrophysics, California Institute of Technology, MC 249-17, 1216 E California Boulevard, Pasadena, CA, 91125, USA}
\email{wynnjg@caltech.edu}

\author[orcid=0000-0002-5619-4938, sname=Kasliwal,gname=Mansi]{M. M. Kasliwal}
\affiliation{Cahill Center for Astrophysics, California Institute of Technology, MC 249-17, 1216 E California Boulevard, Pasadena, CA, 91125, USA}
\email{mansi@astro.caltech.edu}

\author[0000-0001-8985-2493]{Erez A. Zimmerman}
\affiliation{Department of Particle Physics and Astrophysics,Weizmann Institute of Science, 234 Herzl St, 7610001 Rehovot, Israel}
\email{erezimm@gmail.com}

\begin{abstract}
X-ray observations of Type II supernovae (SNe~II) probe the physics of supernova (SN) shocks and the mass-loss histories of their progenitor stars. We present multi-epoch, X-ray observations of SN~II 2024ggi ($D \approx 7.2$~Mpc) from {\it Swift}-XRT, {\it Chandra X-ray Observatory} and {\it XMM-Newton}, which cover $\sim 1 - 344$~days since first light. We analyze these observations using a new open-source Python package called \texttt{XSNAP}, which standardizes a unified command-line interface for instrument-specific reduction and spectral extraction. \texttt{XSNAP} introduces application programming interfaces for per-epoch spectral modeling through \texttt{PyXspec} and \texttt{emcee} Markov chain Monte Carlo fitting. We employ {\tt XSNAP} to model the multi-epoch X-ray spectra of SN~2024ggi with an absorbed thermal bremsstrahlung model and calculate a steady progenitor mass-loss rate of $(6.2\pm0.2)\times10^{-5}\,M_{\odot}\,\mathrm{yr^{-1}}$ $(v_{\rm wind} = 20 \ \rm km \ s^{-1})$, for which the detected X-ray emission traces the final $\sim117$ years before explosion. The software is publicly available on GitHub, with a released package on the Python Package Index (PyPI).
\end{abstract}


\section{Introduction} \label{sec:intro}

X-ray observations of supernovae (SNe) have become increasingly common in the last few decades. Notably, improvements in X-ray spectroscopy allow for the study of element distributions and temperatures in supernova remnants as well as distinguish between thermal (e.g., Bremsstrahlung) and non-thermal (e.g., synchrotron or inverse Compton) emission \citep{2012A&ARv..20...49V}. A recent census from \cite{universe11050161} reports that almost all SNe detected in X-rays are from a core-collapse origin except for SN 2012ca which was classified as Type Ia-CSM by \cite{10.1093/mnras/stu2435}. \cite{universe11050161} suggests the following X-ray characteristics of core-collapse SN subtypes: Type Ib/c SNe are typically non-thermal; Type IIP with low progenitor mass-loss $(\dot{M} \sim 10^{-7} \rm \ M_{\odot} \ yr^{-1})$ are likely non-thermal, with thermal emission becoming dominant as $\dot{M}$ increases; Type IIn events are the most X-ray luminous and clearly thermal; and Type IIb SNe are mostly thermal with a subclass of compact progenitors (cIIb) showing non-thermal emission.

X-ray emission in SNe arises when shocks, created from the fast SN ejecta interacting with the circumstellar medium (CSM), sweep up the CSM, heat the gas to high temperatures, and then the post-shock region cools. In mass coordinates, the resulting shock waves propagate in two directions: inward (reverse shock) and outward (forward shock). Radiation from the reverse shock typically has lower temperatures than that from the forward shock \citep{Chevalier2017}. Moreover, X-rays provide a direct probe of the CSM density profile, enabling estimates of the progenitor’s mass-loss rate in the final years before explosion \citep{1982ApJ...259..302C, 1982ApJ...258..790C, Chevalier2017, universe11050161}. For example, density profiles inferred from X-ray modeling of Type II SNe are often compared to the self-similar solution of \cite{1982ApJ...258..790C}, in which $\rho \propto r^{-s}$ with $s=2$ for a typical Type II SNe. 

Observations over the past two decades have validated and revealed diverse results across SN subtypes. Type Ib/c SNe usually show low mass-loss rates, except some with thermal emission whose mass-loss may reach $\gtrsim 10^{-5}  \rm \ M_{\odot} \ yr^{-1}$, with most extreme cases can be up to $10^{-3}  \rm \ M_{\odot} \ yr^{-1}$ \citep{universe11050161}. Type IIP/L SNe typically exhibit lower mass-loss rates of around $\sim 10^{-7} - 10^{-5}  \rm \ M_{\odot} \ yr^{-1}$, e.g., SN 2013ej \citep{Chakraborti_2016}, SN 2017eaw \citep{Szalai_2019}, and SN 2004et \citep{10.1111/j.1365-2966.2007.12258.x} \citep[although there are exceptions e.g., SN 2023ixf,][]{AJ2025}. Type IIn SNe have the greatest mass-loss rates among SN subtypes, i.e. $\gtrsim 10^{-3} \rm \ M_{\odot} \ yr^{-1}$, e.g. SN 2017hcc \citep{10.1093/mnras/stac2915}, SN 2010jl \citep{Fransson_2014}, SN~2020wyx \citep{Baer25}, or SN~2005ip \citep{2014ApJ...780..184K}. Finally, Type IIb SNe generally have mass-loss rates of around $10^{-5} - 10^{-3} \rm \ M_{\odot} \ yr^{-1}$, e.g., SN 2018gk \citep{10.1093/mnras/stab629}, SN 2013df \citep{2015ApJ...807...35M}, and SN 2008ax \citep{2009ApJ...704L.118R}.

An example of a well-studied Type IIP SN in the X-rays is SN 2023ixf, discovered by K. Itagaki on 2023 May 19 in the M101 galaxy \citep{2023TNSTR1158....1I}. It was first observed in at X-ray wavelengths with the \textit{Neil Gehrels Swift Observatory X-ray Telescope} (\textit{Swift}-XRT) at $\sim 1$~day after first light, which was then complimented with follow-up observations from the \textit{Chandra X-ray Observatory (CXO)}, \textit{XMM-Newton (XMM)}, and \textit{Nuclear Spectroscopic Telescope Array (NuSTAR)} \citep{2025Univ...11..231J}. Multiple analyses have been done with the absorbed plasma model \citep{2024ApJ...963L...4C} and the absorbed Bremsstrahlung model \citep{2023ApJ...952L...3G, 2024Natur.627..759Z, Panjkov24, AJ2025, WJG25c} across the soft and hard X-rays, with the X-ray luminosity of $\sim10^{39} - 10^{40} \ \rm erg \ s^{-1}$ within the first $\sim10^2 \ \rm days$. The findings of these studies are consistent and imply that SN 2023ixf interacted with dense, confined CSM and the red supergiant (RSG) progenitor mass-loss rate was around $10^{-4} - 10^{-3} \ \rm M_{\odot} \ yr^{-1}$, one to two order magnitude higher than the upper-limit of SNe~IIP detected in X-rays \citep{universe11050161}. 

SN 2024ggi, another nearby Type II SN \citep{2024TNSAN.103....1H, 2024TNSCR1031....1Z}, is a well-monitored SN across multiple wavelengths: ($\gamma$-rays \citep{2024ATel16601....1M}, X-rays \citep{2024ATel16588....1Z, 2024ATel16587....1M}, optical bands \citep{2024TNSAN.102....1C, 2024TNSAN.101....1K, 2024TNSAN.108....1K, 2024TNSAN.109....1R, 2025TNSAN..22....1W}, radio frequencies \citep{2024ATel16616....1R}, and centimeter wavelengths \citep{2024ATel16612....1C}). The SN was first discovered by the Asteroid Terrestrial-impact Last Alert System (ATLAS) on 2024-04-11 (MJD 60411.14) \citep{2024TNSTR1020....1T, 2024TNSAN.100....1S, 2025ApJ...983...86C} and was officially classified as a Type IIP \citep{2024ApJ...970L..18Z, 2025ApJ...983...86C}. Previous studies have analyzed UV/optical/IR observations and inferred that the RSG progenitor mass-loss rate was $10^{-3} - 10^{-2} \ \rm M_{\odot} \ yr^{-1}$ \citep{2024ApJ...972L..15S, 2024ApJ...970L..18Z, Jacobson-Galan2024, 2025arXiv250810573A, 2025ApJ...983...86C, 2025A&A...699A..60E}. Additionally, \cite{2024ApJ...969L..15X} estimated the progenitor mass-loss rate is $< 3 \times 10^{-6}  \ \rm M_{\odot} \ yr^{-1}$ from pre-explosion imaging and suggested that SN 2024ggi had an enhancement of mass-loss rate in the century leading up to the explosion. Use of nebular spectroscopy has indicated a progenitor mass in the range of $11 - 15.2 \  \rm M_{\odot}$ \citep{2024ApJ...969L..15X, 2025arXiv250810573A, 2025arXiv250705803D, 2025A&A...699A..60E, Ferrari25, Hueichap25}. In this work, we present and analyze the X-ray observations of SN 2024ggi from \textit{CXO}, \textit{Swift}-XRT, and \textit{XMM}. SN 2024ggi is located at $(\alpha = \rm 11^h18^m22.09^s, \delta=-32^{\circ}50'15.26'')$ in the host galaxy NGC 3621. We adopt a time of first light corresponding to MJD $60410.80 \pm 0.34$ days, redshift of $z=0.002215$ \citep{Jacobson-Galan2024}, and a distance of $7.2 \pm 0.2$ Mpc \citep{Saha_2006}. 

In Section \ref{sec:pipeline}, we describe the design, availability and applications of the ``X-ray Supernova Analysis Pipeline (XSNAP).'' In Section \ref{sec:observation}, we present multi-epoch X-ray observations of SN 2024ggi, and our analysis of SN 2024ggi and its surrounding CSM are described in Section \ref{sec:analysis}. We discuss our findings in Section \ref{sec:discussion} and summarize them in Section \ref{sec:conclusion}.

\section{The Pipeline} \label{sec:pipeline}

Each X-ray observatory, such as {\it CXO}, {\it XMM}, and \textit{Swift}-XRT, and {\it NuSTAR}, employs mission-specific data formats, calibration pipelines, and analysis software. Consequently, this hinders reproducibility in analyzing X-ray observations as astronomers need to work out distinct guidelines for each mission. While there have been past efforts to streamline the process, such as \texttt{YAXX}\footnote{\url{https://cxc.harvard.edu/contrib/yaxx}} (Yet Another X-ray Xtractor; \citeauthor{2006HEAD....9.1358A} \citeyear{2006HEAD....9.1358A}), which supported both {\it CXO} and {\it XMM-Newton} data, these tools are no longer actively maintained. More recent softwares, like \texttt{XGA}\footnote{\url{https://xga.readthedocs.io/en/latest/}} (X-ray: Generate and Analyze; \citeauthor{2022arXiv220201236T} \citeyear{2022arXiv220201236T}) or \texttt{BXA}\footnote{\url{https://johannesbuchner.github.io/BXA/}} (Bayesian X-ray Analysis; \citeauthor{2014A&A...564A.125B} \citeyear{2014A&A...564A.125B}), offer modern functionality but are limited to a single instrument (e.g., {\it XMM}), not designed as end-to-end pipelines from spectral extraction, or not specifically developed for SNe, even though some may still be adaptable for such studies. At present, there is no open-source, Python-based pipeline that unifies spectral extraction and analysis of SN X-ray spectra from {\it CXO}, {\it XMM}, \textit{Swift}-XRT, and {\it NuSTAR}. Consequently, we developed \texttt{XSNAP} (X-ray Supernova Analysis Pipeline), an open-source Python package intended to unify end-to-end X-ray data processes from spectral extraction to CSM analysis. 

\texttt{XSNAP} is organized into two stages: spectrum extraction and spectral modeling and is designed to streamline all the processes described in Sections \ref{sec:observation} and \ref{sec:analysis}, including source detection, deriving count-rates, spectral fitting and deriving unshocked CSM densities. A typical \texttt{XSNAP} workflow consits of (i) generating source and background regions, (ii) extracting and reducing raw spectrum through the Python Command-line Interface (CLI) scripts, and (iii) analyzing spectrum from the Python Application Programming Interface (API) modules. This unified workflow is applied to the analysis of SN 2024ggi presented in Sections \ref{sec:observation} and \ref{sec:analysis}. The spectrum extraction component standardizes the procedures for observations from \textit{CXO}, \textit{XMM}, and \textit{Swift}-XRT (Section \ref{sec:observation}), implementing them as dedicated Python CLI scripts. In addition, a script is made seperately for observations from {\it NuSTAR}; we follow the standard guide from NuSTAR Data Analysis Software (\texttt{NuSTARDAS}\footnote{\url{https://heasarc.gsfc.nasa.gov/docs/nustar/analysis/nustar_quickstart_guide.pdf}}) \citep{Harrison_2013} inside \texttt{HEASoft} \citep{2014ascl.soft08004N}. \texttt{XSNAP} also provides a script for generating source and background regions, making use of \texttt{ximage} and (optionally) \texttt{SAOImage DS9} \citep{2003ASPC..295..489J}. Similarly, we developed the spectrum and CSM analysis sector by wrapping all the processes from Section \ref{sec:analysis} to each separate Python API module. We then packaged the pipeline in a Python Package Index (PyPI) environment and published it to Github\footnote{\url{https://github.com/fercananything/XSNAP/}}, PyPI\footnote{\url{https://pypi.org/p/xsnap/}}, and its own documentation website\footnote{\url{https://xsnap.org/}}.

\section{Observations} \label{sec:observation}

{
\setlength{\tabcolsep}{12pt}
\begin{deluxetable*}{l c c c c c c}
\tablecaption{Best‐fitting Parameters of the Absorbed Thermal Bremsstrahlung Model for SN2024ggi\label{tab:fit_SN2024ggi}}
\tablehead{
  \colhead{Phase} &
  \colhead{Instrument} &
  \colhead{$\mathrm{N}_{\rm H, \ int}$} &
  \colhead{$T$\tablenotemark{a}} &
  \colhead{$\log_{10}$(Flux)\tablenotemark{b}} &
  \colhead{$\log_{10}$(Flux)\tablenotemark{b}} &
  \colhead{Norm\tablenotemark{c}} \\ [-6pt]
  \colhead{(days)} &
  \colhead{} &
  \colhead{$\left(10^{22} \ \rm cm^{-2}\right)$} &
  \colhead{(keV)} &
  \colhead{(Absorbed)} &
  \colhead{(Unabsorbed)} &
  \colhead{($10^{-4} \ \rm \mathbf{cm^{-5}}$)}
}
\startdata
  3.33 &	Swift-XRT &	$7.36^{+3.93}_{-2.89}$ &	47.81 &	$-12.57^{+0.13}_{-0.14}$ &	$-12.30^{+0.13}_{-0.14}$ &	$0.96^{+0.41}_{-0.30}$ \\
  8.36	& Swift-XRT	& $9.35^{+6.33}_{-3.92}$ &	37.97 &	$-12.54^{+0.12}_{-0.12}$ &	$-12.23^{+0.12}_{-0.12}$ &	$1.09^{+0.56}_{-0.37}$ \\
 10.90 & CXO       & $2.83^{+0.57}_{-0.51}$ & 35.54 & $-12.51^{+0.04}_{-0.04}$ & $-12.32^{+0.04}_{-0.04}$ & $0.88^{+0.10}_{-0.09}$ \\
 16.20 & CXO       & $1.33^{+0.42}_{-0.38}$ & 32.18 & $-12.69^{+0.05}_{-0.04}$ & $-12.56^{+0.05}_{-0.04}$ & $0.51^{+0.07}_{-0.06}$ \\
 24.49 & Swift‐XRT & $0.06^{+0.53}_{-0.06}$ & 29.02 & $-12.81^{+0.12}_{-0.17}$ & $-12.77^{+0.12}_{-0.17}$ & $0.31^{+0.13}_{-0.08}$ \\
 34.76 & Swift‐XRT & $0.00^{+0.37}_{-0.00}$ & 26.59 & $-12.87^{+0.14}_{-0.16}$ & $-12.84^{+0.14}_{-0.16}$ & $0.26^{+0.11}_{-0.08}$ \\
 43.20 & Swift‐XRT & $0.00^{+0.08}_{-0.00}$ & 25.18 & $-12.96^{+0.13}_{-0.18}$ & $-12.94^{+0.13}_{-0.18}$ & $0.21^{+0.07}_{-0.05}$ \\
 55.03 & XMM       & $0.00^{+0.02}_{-0.00}$ & 23.7 & $-13.10^{+0.06}_{-0.08}$ & $-13.07^{+0.06}_{-0.08}$ & $0.15^{+0.02}_{-0.02}$ \\
 85.40 & XMM       & $0.01^{+0.05}_{-0.01}$ & 21.24 & $-13.25^{+0.04}_{-0.05}$ & $-13.22^{+0.04}_{-0.05}$ & $0.11^{+0.01}_{-0.01}$ \\
\enddata
\tablecomments{All errors are reported at $1\sigma$ c.l.}
\tablenotetext{a}{Value is fixed in model fit.}
\tablenotetext{b}{$\log_{10}$(Flux) is from 0.3–10 keV. Flux units in $\rm erg \ cm^{-2} \ s^{-1}$}
\tablenotetext{c}{Normalization of the Bremsstrahlung model defined as $\rm Norm \equiv \frac{3.02 \times 10^{-15}}{4 \pi d^2} \int n_e n_I \ dV$, where $n_e$ and $n_I$ are the electron and ion number densities in $\rm cm^{-3}$ and $d$ is the distance to the source in $\rm \mathbf cm$.}
\end{deluxetable*}
}

\begin{figure*}[htb!]
  \gridline{
    \fig{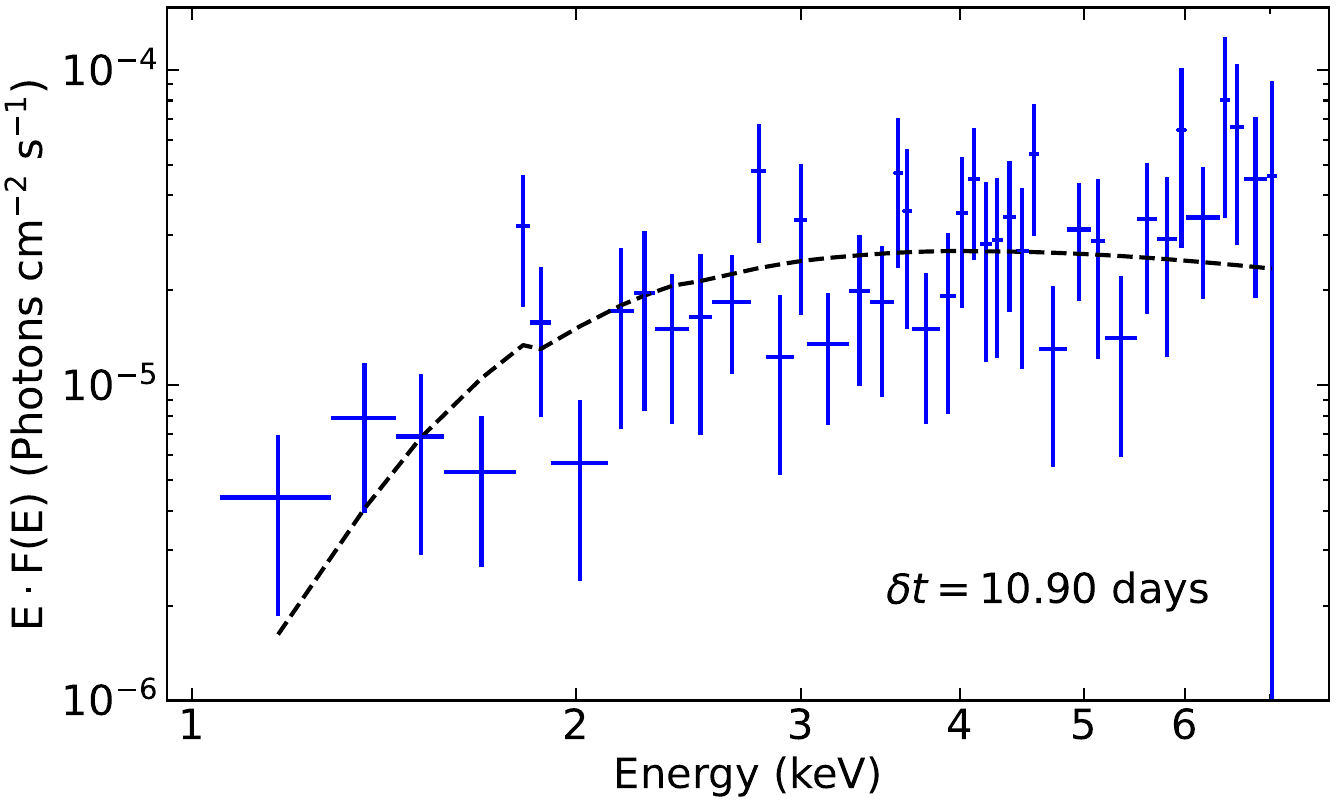}{0.5\textwidth}{}
    \fig{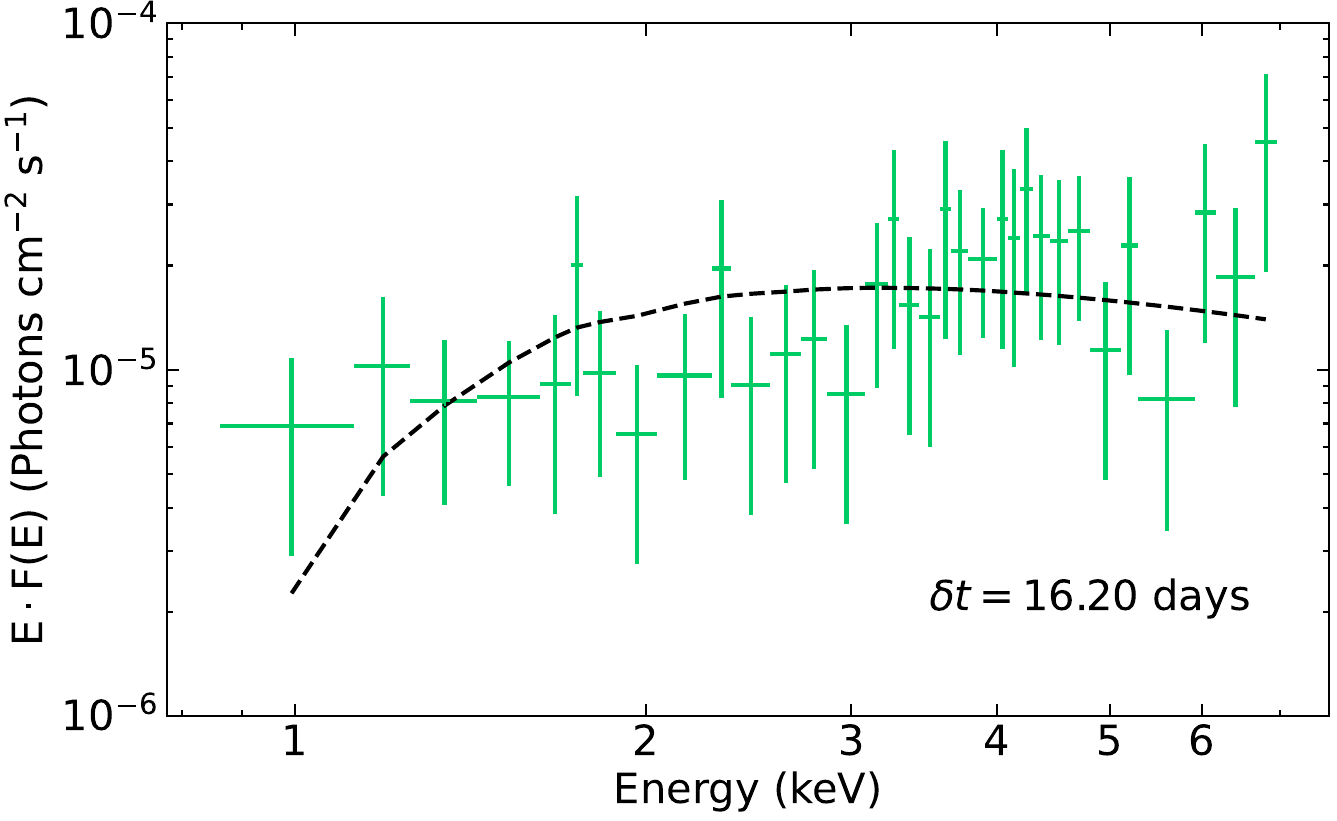}{0.5\textwidth}{}
  }
  \gridline{
    \fig{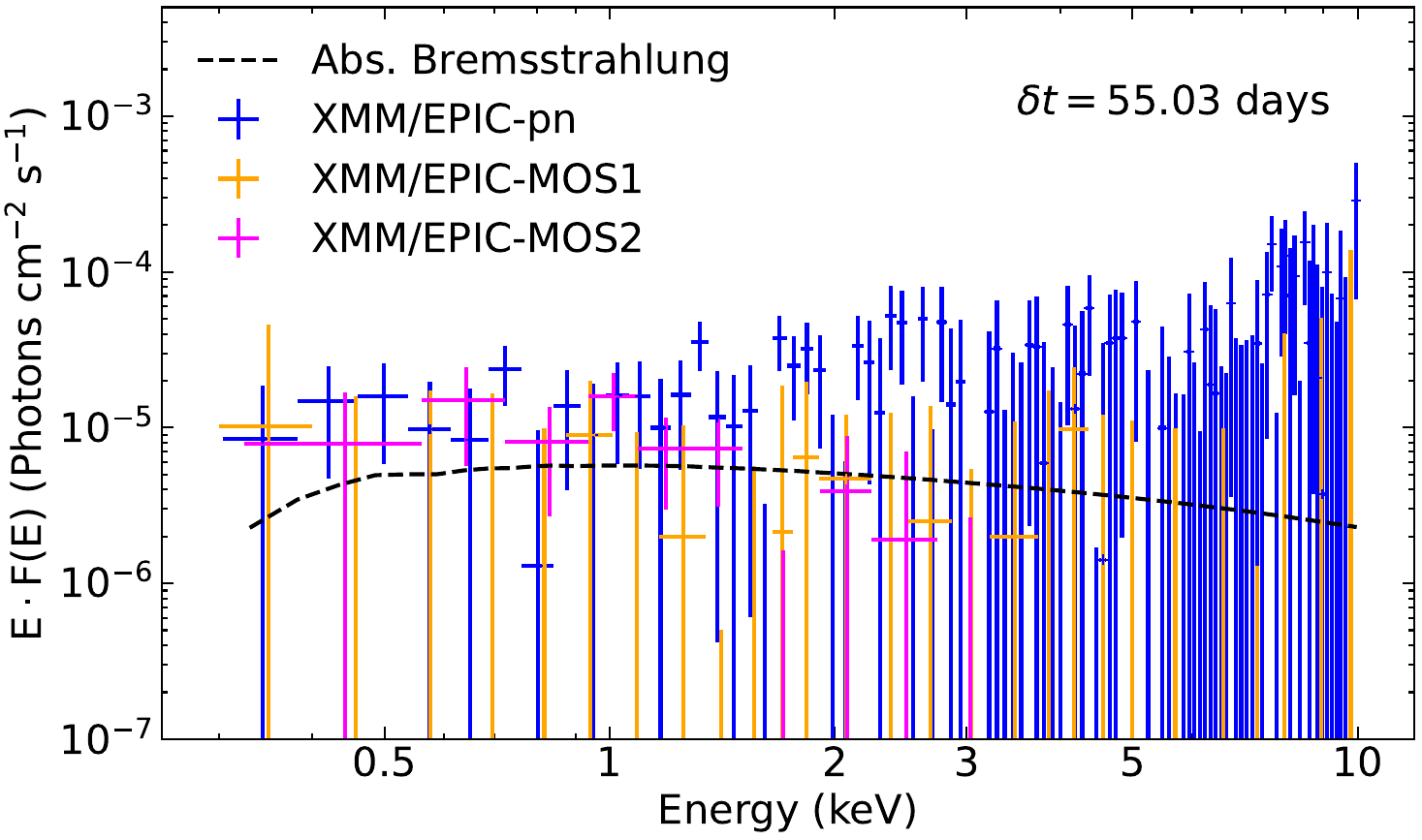}{0.5\textwidth}{}
    \fig{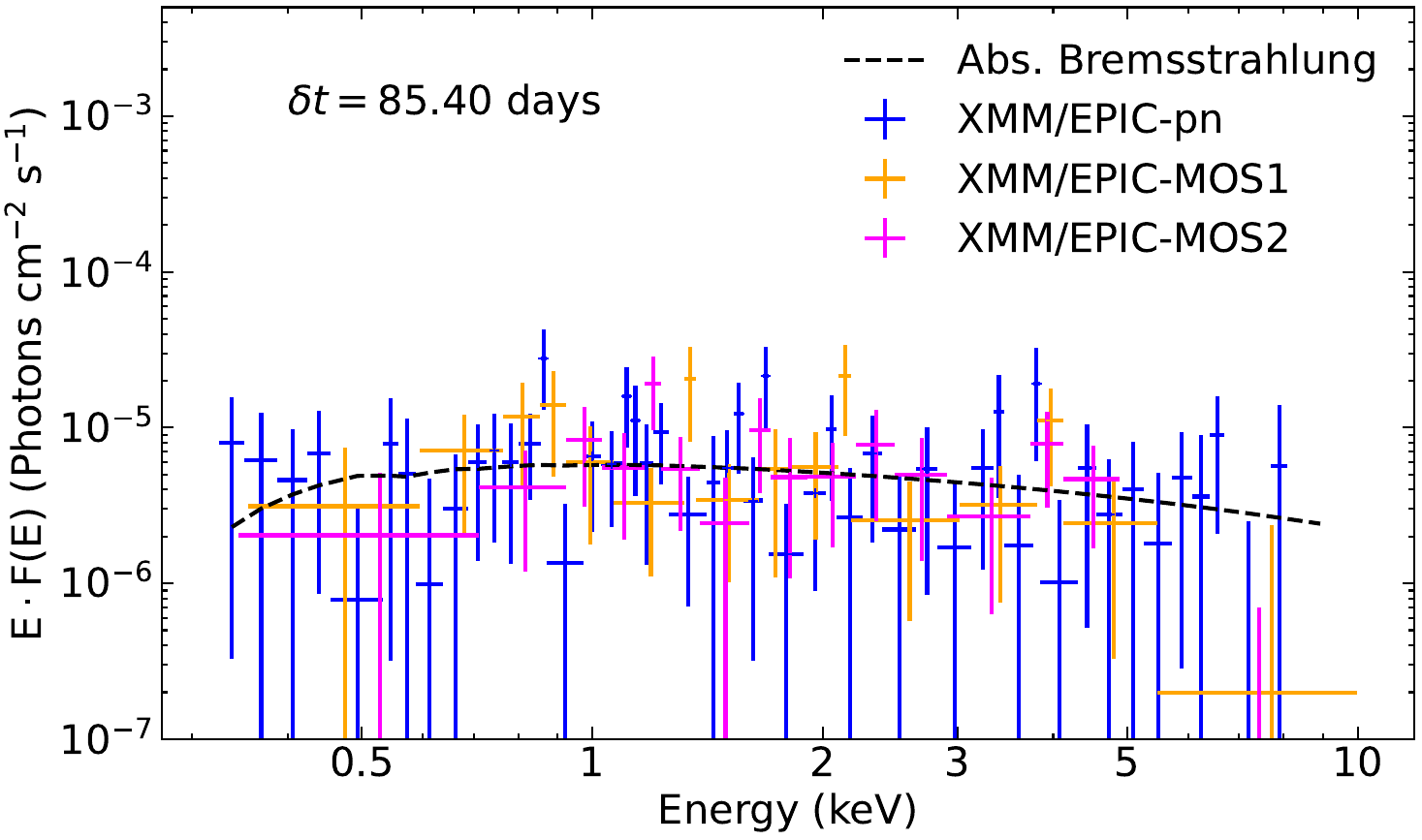}{0.5\textwidth}{}
  }
  
  \caption{\textit{Top:} Plot of fitted spectrum from {\it CXO} at $\delta t=10.90 \ \rm days$ (\textit{left}) and $\delta t=16.20 \ \rm days$ (\textit{right)} since first light. Blue (\textit{left}) and green (\textit{right}) points are observed X-ray spectra and black dashed lines are the best-fit absorbed thermal Bremsstrahlung model. \textit{Bottom:} Plot of fitted spectrum from {\it XMM} data $\delta t=55.03 \ \rm days$ (\textit{left}) and $\delta t=85.40 \ \rm days$ (\textit{right)} since first light. Observed spectra are represented in blue (EPIC-pn), orange (EPIC-MOS1), and magenta (EPIC-MOS2) and black dashed lines are the best-fit absorbed thermal Bremsstrahlung model.}
  \label{fig:spectrum}
\end{figure*}

\begin{figure*}[htb!]
  \gridline{
    \fig{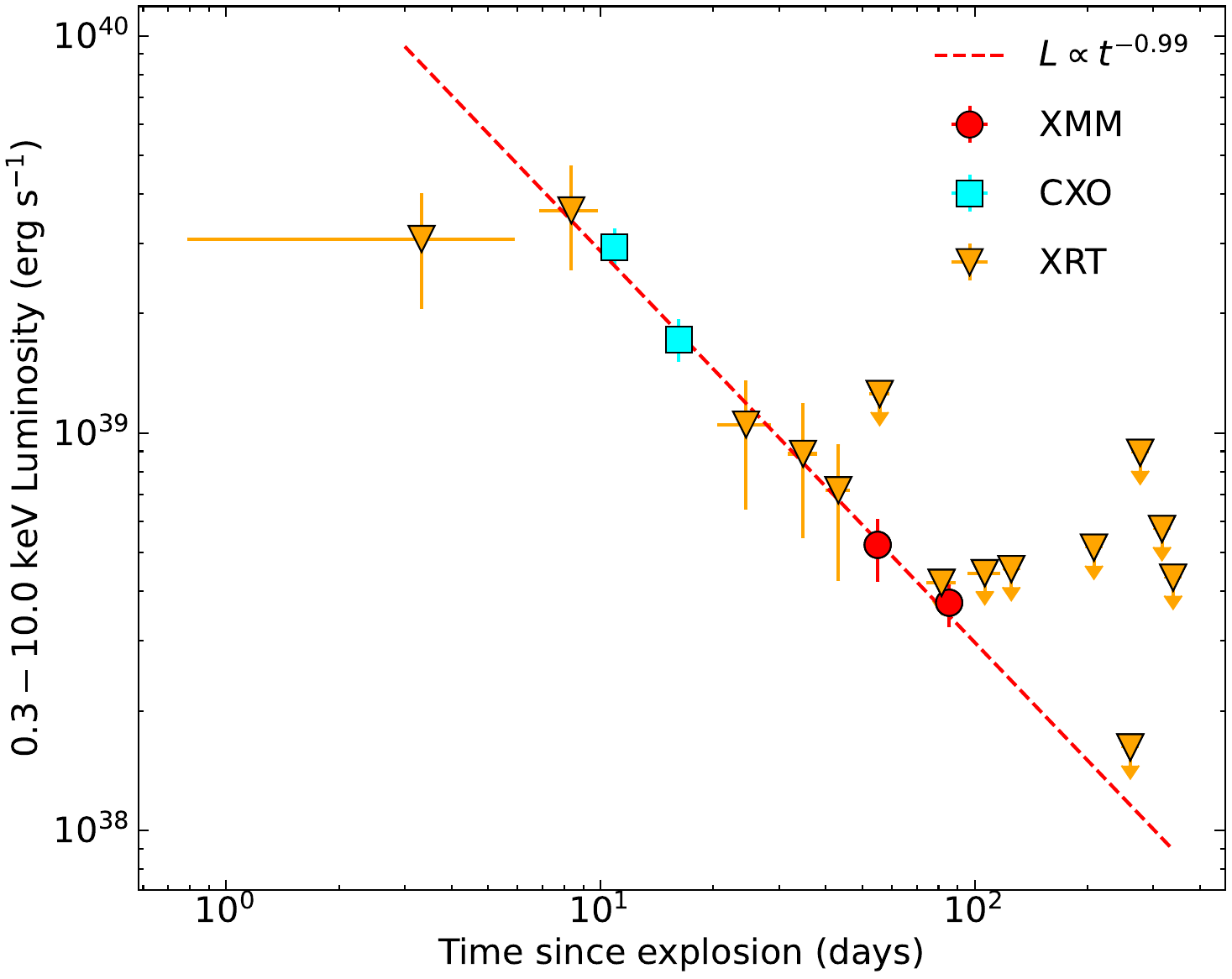}{0.5\textwidth}{}
    \fig{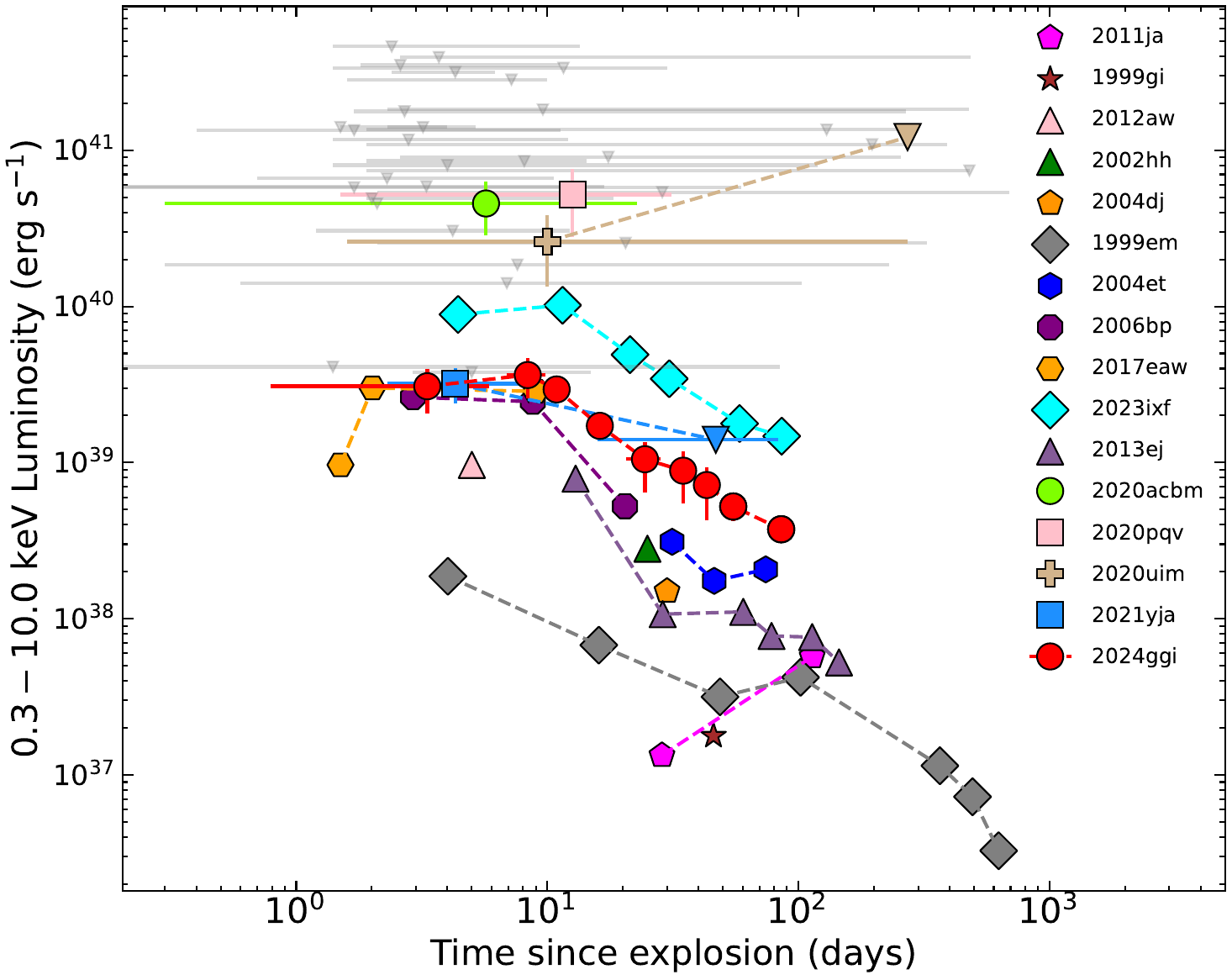}{0.5\textwidth}{}
  }
  
  \caption{\textit{Left}: Plot of fitted unabsorbed $0.3 - 10 \ \rm keV$ X-ray luminosity light curve from {\it XMM} (red circles), CXO (cyan squares), and {\it Swift}-XRT (orange triangles) observations. The red dashed line is the fitted line with $L \propto t^{-0.99}$. \textit{Right}: Plot of unabsorbed $0.3 - 10 \ \rm keV$ X-ray luminosity light curve of SN 2024ggi compared to a sample of Type IIP SNe. References: \citet{Schlegel_1999, Schlegel_2001}, \cite{Pooley_2002}, \cite{2002IAUC.8024....2P, 2004IAUC.8390....1P}, \cite{10.1111/j.1365-2966.2007.12258.x}, \cite{Immler_2007}, \cite{2012ATel.3995....1I},  \cite{Chakraborti_2013, Chakraborti_2016}, \cite{Szalai_2019}, \cite{2024ApJ...970...96I}, \cite{AJ2025}}
  \label{fig:lumin}
\end{figure*}

\begin{figure*}[hbt!]
  \gridline{
    \fig{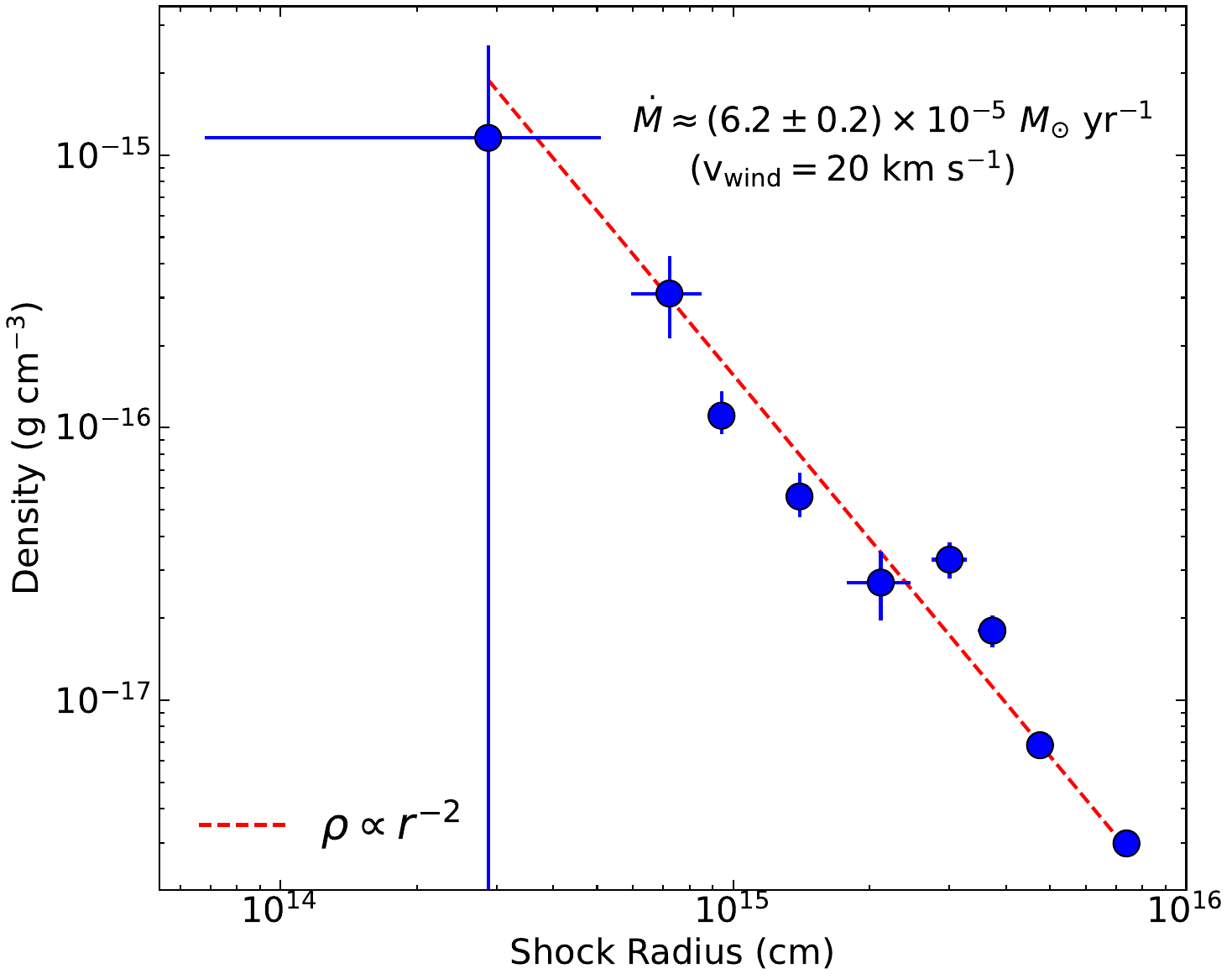}{0.5\textwidth}{}
    \fig{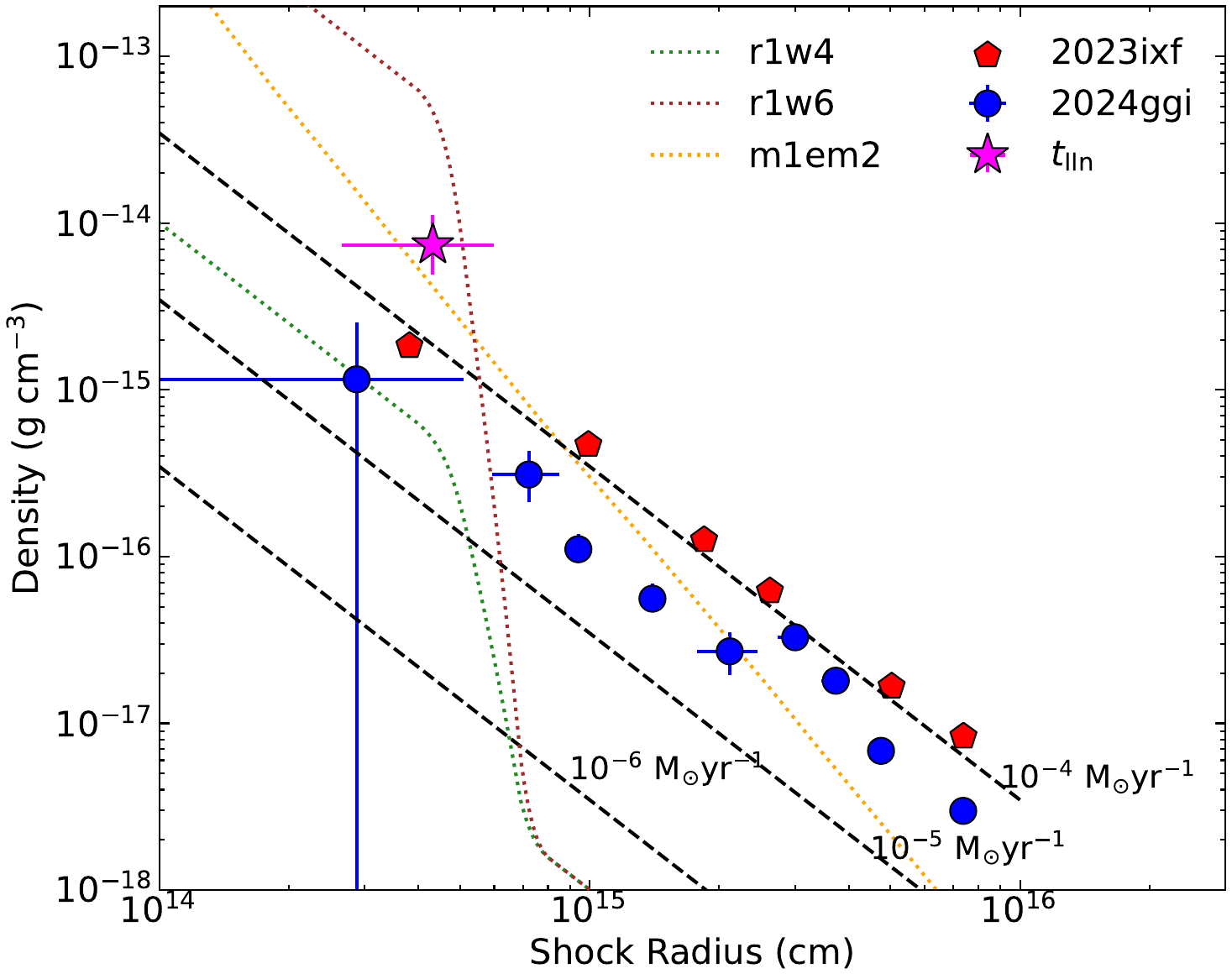}{0.5\textwidth}{}
  }
  
  \caption{\textit{Left}: Plot of fitted density profile. The blue circles are the data derived from analysis and the red dashed line is the fitted line with $\rho \propto r^{-2}$. \textit{Right}: Plot of fitted density profile to a previous study by \cite{Jacobson-Galan2024}. The r1w4 (green dashed line), r1w6 (red dashed line), and m1em2 (orange dashed line) are best-matched models from \texttt{CMFGEN} model grid  \citep{Jacobson-Galan2024, Jacobson-Galan2024b}. The magenta star is the transition point where the CSM goes from optically thick to thin to electron scattering \citep{Jacobson-Galan2024}. The blue circles are the data derived from analysis and the blue dashed line is the fitted line with $\rho \propto r^{-2}$. The purple pentagons are SN 2023ixf densities derived from X-ray observations \citep{AJ2025}. The grey dashed lines are reference density profile lines if the mass-loss rates are $10^{-4} \ \rm M_{\odot} \ yr^{-1}$, $10^{-5} \ \rm M_{\odot} \ yr^{-1}$, and $10^{-6} \ \rm M_{\odot} \ yr^{-1}$ respectively for $v_{\rm wind} = 20 \ \rm km \ s^{-1}$.}
  \label{fig:density}
\end{figure*}

\subsection{Chandra X-ray Observatory}
The {\it CXO} ACIS-S observed SN 2024ggi in two epochs: one on 2024 April 21 (ObsID: 29383; $\delta t = 10.90 \ \rm days$; PI: E. Zimmerman) with an exposure time of 14.88 ks and the other on 2024 April 26 (ObsID: 29384; $\delta t = 16.20 \ \rm days$; PI: E. Zimmerman) with an exposure time of 14.58 ks. Details of these two observations are reported in Table \ref{tab:obslogggi}. 

We reduced both observations using \texttt{CIAO v4.17} \citep{2006SPIE.6270E..1VF} and their corresponding calibration files. Source detection and estimating source-free background regions were utilized through \texttt{ximage v4.5.1} from \texttt{HEASoft} with signal-to-noise ratio threshold of three $(\rm SNR_{thresh}=3)$ and human vetting with \texttt{SAOImage DS9}. For each epoch, we extracted the spectrum via \texttt{specextract} and used a circular source region with a radius of $2''$ and circular background region with a radius of $45''$. 

\subsection{XMM-Newton}
Similar to {\it CXO}, {\it XMM} observed SN 2024ggi in two epochs: one on 2024 June 04 (ObsID: 0882480901; $\delta t = 55.03 \ \rm days$; PI: W. V. Jacobson-Gálan) and the other on 2024 July 05 (ObsID: 0882481001; $\delta t = 85.40 \ \rm days$; PI: W. V. Jacobson-Gálan). Details of these two observations are reported in Table \ref{tab:obslogggi}. 

We reduced both observations from the three instruments of {\it XMM}, i.e. European Photon Imaging Cameras (EPIC)-pn, MOS1, and MOS2, using the \texttt{Scientific Analysis System (SAS) v22.1.0} \citep{2014ascl.soft04004S} and corresponding calibration files. Source detection and estimating source-free background regions were utilized through \texttt{ximage v4.5.1} from \texttt{HEASoft} with signal-to-noise ratio threshold of three $(\rm SNR_{thresh}=3)$ and human vetting with \texttt{SAOImage DS9}. For each epoch, we extracted the spectrum via \texttt{evselect} and used a circular source region with a radius of $25''$ and circular background region with a radius of $125''$. 

\subsection{Swift X-ray Telescope}
The \textit{Swift}-XRT \citep{2005SSRv..120..165B} onboard the {\it Neil Gehrels Swift Observatory} \citep{2004ApJ...611.1005G} started observing SN 2024ggi on 2024 April 11 UT 14:04:10 $(\delta t = 0.79 \ \rm days)$. In this work, we analyzed \textit{Swift}-XRT datasets collected until 2025 March 20.  As each observation has relatively small exposure time $(\sim 1.5 \ \rm ks)$, we binned a few epochs together, ranging from 10-day to 30-day intervals until the source was detected. However, SN~2024ggi was not detected in \textit{Swift}-XRT images after 2024 June 02 $(\delta t = 52.20 \ \rm days)$ and these observations were analyzed as upper-limits (see Section \ref{sec:analysis}). Details of each stacked observation is reported in Table \ref{tab:obslogggi}. 

We reduced each stacked observation using \texttt{HEASoft v6.35}, the Swift XRT Data Analysis Software (\texttt{SWXRTDAS}; version 3.7.0), and the Swift Calibration Database (CALDB) version 20250609. Source detection and estimating source-free background regions were utilized through \texttt{ximage v4.5.1} from \texttt{HEASoft} with signal-to-noise ratio threshold of three $(\rm SNR_{thresh}=3)$ and human vetting with \texttt{SAOImage DS9}. For each epoch, we extracted the spectrum via \texttt{xrtpipeline}, followed the standard practice from \citet{2009MNRAS.397.1177E, 10.1093/mnras/sts066}, and used a circular source region with a radius of $25''$ and circular background region with a radius of $125''$.

\section{Analysis} \label{sec:analysis}

\subsection{X-ray Spectral Modeling}

We modeled and fit each observation in Table \ref{tab:obslogggi} with \texttt{PyXspec} \citep{2021ascl.soft01014G}, the Python interface to \texttt{XSPEC} \citep{1996ASPC..101...17A}. We fit them separately except for {\it XMM} data where we did a joint fit for all three instruments, EPIC-pn, MOS1, and MOS2. In this work, we adopted the solar abundances from \cite{2009ARA&A..47..481A}, i.e., \texttt{abund aspl} within \texttt{PyXspec}. We set the Galactic neutral hydrogen ($\rm N_H$) column density in the direction of SN2024ggi of $\mathrm{N}_{\mathrm{H, \ MW}} = 6.59 \cdot 10^{20} \ \mathrm{cm}^{-2}$ \citep{2016A&A...594A.116H}. For the spectral fits, we restricted the {\it Chandra}/ACIS data to the $1 - 8 \ \mathrm{keV}$ energy range, while the {\it Swift}-XRT and {\it XMM} spectra were fit in the $0.3 -10 \ \mathrm{keV}$ band.

We then fit these data using an absorbed thermal Bremsstrahlung model (\texttt{tbabs*ztbabs*bremss}), but were unable to constrain the model temperature. Therefore, we estimated and fixed the temperatures of each observation using the self-similar solution $T = 34 \ \rm keV \ (t/13d)^{-0.25}$ \citep{2024ApJ...963L...4C}. This temperature evolution follows the self-similar solutions by \cite{1982ApJ...258..790C, Chevalier2017} for a standard wind with a constant mass-loss rate, $\rho \propto r^{-2}$. For such a case, the post-shock temperature evolves as $T \propto t^{-2/(n-2)}$, which becomes $T \propto t^{-0.25}$ after assuming $n=10$. We adopted a normalization of $T = 34~\rm keV$ at $t = 13~\rm d$, as utilized for SN~2023ixf by \cite{2024ApJ...963L...4C}.  Finally, we  modeled the intrinsic $\rm N_H$ and Bremsstrahlung normalization and estimated the absorbed and unabsorbed fluxes in $0.3 - 10 \ \rm keV$. We also repeated this analysis while adopting the temperature evolution of SN~2023ixf, $T \propto t^{-0.5}$ \citep{AJ2025}, which yielded consistent results to those found with a self-similar solution temperature decline rate. Based on the resulting best‐fit parameters, the X-ray spectra of SN 2024ggi reveal SN ejecta interaction with dense CSM: (i) the emission was highly absorbed at early times, (ii) the absorption decreased with time, and (iii) of thermal nature, consistent with previous studies \citep{2017ApJ...835..140M, 2022ApJ...930...57T}. The best-fit parameters are reported in Table \ref{tab:fit_SN2024ggi} and contour plots of the first four epochs data from {\it Swift}-XRT and {\it CXO} are shown in \ref{fig:contour}. In addition, the fitted spectrum for {\it CXO} and {\it XMM} observations are shown in Figure \ref{fig:spectrum}.

We classified an observation as a non-detection when the source, SN~2024ggi, has a signal-to-noise ratio below three $(\rm SNR<3)$. For those observations, we used the \texttt{FakeIt} command from \texttt{PyXspec} to simulate fake spectra with the same absorbed thermal Bremsstrahlung model (\texttt{tbabs*ztbabs*bremss}). In the model, we set the intrinsic $\rm N_H$ for non-detections to be $\rm N_{H, \ int} = 0$ and fixed the temperature to follow $T = 34 \ \rm keV \ (t/13d)^{-0.25}$. We then used the fake spectrum flux and the count rate ratio to estimate the flux upper-limit in $0.3 - 10 \ \rm keV$ band. However, we did not include parameters derived from non-detections for fitting the luminosity decay and the CSM density profile analysis.

We converted all unabsorbed fluxes to luminosities and fit a power-law with \texttt{emcee} Markov chain Monte Carlo (MCMC). We found that the intrinsic $0.3 - 10 \ \rm keV$ luminosity has a decay of $L_X \propto t^{-0.99}$ at $\delta t \gtrsim 8.36$~days. Additionally, in Figure \ref{fig:lumin}, we compare the X-ray light curve to a sample of SNe~II compiled by \cite{2024ApJ...970...96I, AJ2025}.

\subsection{CSM Density Profile Analysis}

We estimated the CSM density profile, $\rho_{\rm CSM}(r)$, by using the normalization of thermal bremsstrahlung model as it is directly proportional to the emission measure (EM). EM is described by:
\begin{equation}
EM \equiv \int n_e n_I \ dV
\end{equation}
where $n_e$ and $n_I$ are the electron and ion number densities in the emitting volume $V$. For a thermal bremsstrahlung model in {\tt Xspec}, the spectral normalization is defined by:
\begin{equation}
\rm Bremss_{Norm} \equiv \frac{3.02 \times 10^{-15}}{4 \pi d^2} EM
\end{equation}
where $n_e$ and $n_I$ are in $\rm cm^{-3}$ and $d$ is the distance to the source in $\rm \mathbf cm$. 
We follow the procedure of \cite{Brethauer_2022} to calculate the unshocked CSM density profile, which goes as:
\begin{equation}
	\rho_{\rm CSM}(r) = \frac{m_p}{4} \left(\frac{2 \times \rm EM(r)\mu_e \mu_I}{V_{\rm FS}(r)}\right)^{1/2}
\end{equation}
where $\mu_e$ and $\mu_I$ are the mean molecular weight of electron and ion respectively; $m_p$ is the proton mass; $ \mathrm{V_{FS}} = \frac{4\pi}{3}f \left(R_{\rm out}^3 - R_{\rm in}^3\right)$ is the emitting forward shock (FS) volume, with $R_{\rm in}$ and $R_{\rm out}$ are the inner and outer radius of the shell respectively and $f$ is the filling factor. In this work, we adopted $R_{\rm out} = 1.2 R_{\rm in}$, $f=1$, $\mu_e=1.14$, and $\mu_I = 1.24$ (assuming solar values).  We also estimated the shock radius by $R_{\rm shock} = v_{\rm shock} \times t$, where $t$ is time since explosion and assumed an average FS velocity $v_{\rm shock} = 10^4 \ \rm km \ s^{-1}$. The mass-loss rate was then estimated by fitting the density profile to $\rho_{CSM}(r) = \frac{\dot{M}}{4 \pi r^2 v_{\rm wind}}$ with the \texttt{emcee} package. Incorporating $v_{\rm wind} = 20 \ \rm km \ s^{-1}$, we inferred a mass-loss rate of $\dot{M} = (6.2 \pm 0.2) \times 10^{-5} \ \rm M_{\odot} \ yr^{-1}$ from the detected emission for the last $\sim117$ years before explosion, which lies between the upper-limit of the mass-loss rates of Type IIP SNe $\sim 10^{-5} \ \rm M_{\odot} \ yr^{-1}$ and the lower-limit of the mass-loss rates of Type IIn SNe $\sim 10^{-3} \ \rm M_{\odot} \ yr^{-1}$ \citep{universe11050161}. This lookback time can be treated as an approximation as they heavily rely on the wind and shock velocities, which both are assumed. The density profile and its comparison to SN 2023ixf and density profile models by \cite{Jacobson-Galan2024} are shown in Figure \ref{fig:density}. 

\section{Discussion} \label{sec:discussion}

Detected X-ray observations of SN 2024ggi up to $\sim 85$ days after explosion constrain both the shock properties and the progenitor activity before core collapse. SN 2024ggi has a progenitor mass-loss rate of $(6.2 \pm 0.2) \times 10^{-5} \ \rm M_{\odot} \ yr^{-1}$ for a steady-state wind velocity of $v_{\rm w} = 20 \ \rm km \ s ^{-1}$ for the last $\sim117$ years before first light based on the phases of detected X-ray emission. This mass-loss rate is higher than what is estimated for most SNe~IIP with X-ray observations (e.g., see \citealt{2024ApJ...970...96I, universe11050161}), but comparable to the SN 2023ixf progenitor mass-loss rate of $\sim 10^{-4} \ \rm \ M_{\odot} \ yr^{-1}$ \citep{AJ2025,WJG25c,Bostroem25} at shock radii $>10^{15}$~cm. Based on Figure 6 in \cite{2024OJAp....7E..47F}, $6.2 \times 10^{-5} \ \rm M_{\odot} \ yr^{-1}$ corresponds to a RSG progenitor mass of $\sim 20 \  \rm M_{\odot}$. This estimate is higher than masses typically inferred from pre-explosion imaging and nebular-phase spectroscopy for CSM-interacting SNe~II (e.g., \citealt{Kilpatrick25, WJG25b}), including SN~2024ggi \citep{Ferrari25, 2024ApJ...969L..15X}. Additionally, the X-ray luminosity peaks around the first $\sim 5$ days after first light and decays as $L_X \propto t^{-0.99}$, which is approximately aligned with a steady-state wind mass-loss who has a luminosity decay of $L \propto t^{-1}$.

A temporal evolution of the intrinsic neutral hydrogen column $(\rm N_{\rm H, \  int})$ provides evidence for a dense, confined CSM (see Table \ref{tab:fit_SN2024ggi}). Although our results agree with \citet{Jacobson-Galan2024} that a dense, confined CSM is present, the densities inferred from the optical spectra are higher than those derived from the X-rays at $<5\times 10^{14}$~cm. This discrepancy, also observed in SN 2023ixf, may arise from CSM asymmetry or clumping \citep{2024ApJ...975..132S, AJ2025, 2025arXiv250503975V}. At early times $(\delta t \lesssim 16 \ \rm days)$, we found $\rm N_{\rm H, \ int} > 10^{22} \ \rm cm^{-2}$, indicating significant X-ray absorption in a dense medium extending up to $r \lesssim 10^{15} \ \rm cm$, before this material overtaken by the SN shock at $\delta t \gtrsim 11 \ \rm days$. All of these results were produced with our pipeline, \texttt{XSNAP}, which streamlines the entire workflow from raw counts to CSM analysis. Beyond this application to SN 2024ggi, \texttt{XSNAP} is broadly applicable: its data reduction and spectral modeling modules can be used for other X-ray sources, while its CSM analysis is specialized for SNe.

\section{Conclusion} \label{sec:conclusion}

Using our \texttt{XSNAP} pipeline, we present our X-ray observation and analysis of SN 2024ggi, constraining both the shock properties and the mass loss rate of its RSG progenitor. In the following, we summarize our main findings of SN 2024ggi:
\begin{itemize}
	\item We estimate a progenitor mass-loss rate of $(6.2 \pm 0.2) \times 10^{-5} \ \rm M_{\odot} \ yr^{-1}$ from the detected emission for $\sim 117$ years before explosion $(v_{\rm wind} = 20 \ \rm km \ s^{-1})$.
	\item SN 2024ggi shows evidence for a dense CSM $(\rm N_{\rm H, \  int} > 10^{22} \ \rm cm^{-2})$ up to $\sim 16$ days after explosion, confined within $r \lesssim 1.4 \times 10^{15} \ \rm cm$. 
	\item SN 2024ggi has a luminosity decline of $L_X \propto t^{-0.99}$ and density profile $\rho \propto r^{-2}$, consistent with SN ejecta interaction in a steady-state, wind-like CSM.
	\item All results were obtained with our new pipeline, \texttt{XSNAP}, demonstrating reproducibility from raw data reduction to CSM analysis. While specialized for supernovae, its reduction and spectral modeling modules are general and applicable to other X-ray sources. \texttt{XSNAP} is publicly available on GitHub and PyPI for community use.
\end{itemize}

SN 2024ggi offers a unique window into the study of SNe~II and the characterization of presence of dense, confined CSM around their progenitor stars. Moreover, \texttt{XSNAP} enables a consistent framework to extend such studies to future X-ray--detected SNe.

\begin{acknowledgements}

    F. acknowledges support from the Caltech Summer Undergraduate Research Fellowship (SURF) program. W.J.-G.\ is supported by NASA through Hubble Fellowship grant HSTHF2-51558.001-A awarded by the Space Telescope Science Institute, which is operated for NASA by the Association of Universities for Research in Astronomy, Inc., under contract NAS5-26555.
    This research has made use of data and/or software provided by the High Energy Astrophysics Science Archive Research Center (HEASARC), which is a service of the Astrophysics Science Division at NASA/GSFC. This paper employs a list of Chandra datasets, obtained by the Chandra X-ray Observatory, contained in the Chandra Data Collection ~\dataset[DOI: 10.25574/cdc.531]{https://doi.org/10.25574/cdc.531}.
\end{acknowledgements}

\facilities{CXO, Swift (XRT), XMM-Newton}
\software{\texttt{Astropy} \citep{astropy:2013, astropy:2018, astropy:2022}, \texttt{CIAO} \citep{2006SPIE.6270E..1VF}, \texttt{corner} \citep{corner}, \texttt{emcee} \citep{2013PASP..125..306F}, \texttt{HEASoft} \citep{2014ascl.soft08004N}, \texttt{NumPy} \citep{harris2020array}, \texttt{Pandas} \citep{mckinney-proc-scipy-2010, reback2020pandas}, \texttt{PyXspec} \citep{2021ascl.soft01014G}, \texttt{SAOImage DS9} \citep{2003ASPC..295..489J}, \texttt{SAS} \citep{2014ascl.soft04004S}, \texttt{SciPy} \citep{2020SciPy-NMeth}, \texttt{XSPEC} \citep{1996ASPC..101...17A}}

\bibliography{main}{}

\begin{thebibliography}{}
\expandafter\ifx\csname natexlab\endcsname\relax\def\natexlab#1{#1}\fi
\providecommand{\url}[1]{\href{#1}{#1}}
\providecommand{\dodoi}[1]{doi:~\href{http://doi.org/#1}{\nolinkurl{#1}}}
\providecommand{\doeprint}[1]{\href{http://ascl.net/#1}{\nolinkurl{http://ascl.net/#1}}}
\providecommand{\doarXiv}[1]{\href{https://arxiv.org/abs/#1}{\nolinkurl{https://arxiv.org/abs/#1}}}

\bibitem[{T. {Aldcroft}(2006){Aldcroft}}]{2006HEAD....9.1358A}
{Aldcroft}, T. 2006, in AAS/High Energy Astrophysics Division, Vol.~9, AAS/High
  Energy Astrophysics Division \#9, 13.58

\bibitem[{K.~A. {Arnaud}(1996){Arnaud}}]{1996ASPC..101...17A}
{Arnaud}, K.~A. 1996, in Astronomical Society of the Pacific Conference Series,
  Vol. 101, Astronomical Data Analysis Software and Systems V, ed. G.~H.
  {Jacoby} \& J.~{Barnes}, 17

\bibitem[{A. {Aryan} {et~al.}(2025){Aryan}, {Higgins}, {Nicholl}, {Chen}, \&
  {Liu}}]{2025arXiv250810573A}
{Aryan}, A., {Higgins}, E., {Nicholl}, M., {Chen}, T.-W., \& {Liu}, Y.-H. 2025,
  \bibinfo{title}{{Constraints on the progenitor and explosion of SN 2024ggi in
  harmony with pre-explosion detection and hydrodynamic simulations},} arXiv
  e-prints, arXiv:2508.10573, \dodoi{10.48550/arXiv.2508.10573}

\bibitem[{M. {Asplund} {et~al.}(2009){Asplund}, {Grevesse}, {Sauval}, \&
  {Scott}}]{2009ARA&A..47..481A}
{Asplund}, M., {Grevesse}, N., {Sauval}, A.~J., \& {Scott}, P. 2009,
  \bibinfo{title}{{The Chemical Composition of the Sun},} \araa, 47, 481,
  \dodoi{10.1146/annurev.astro.46.060407.145222}

\bibitem[{ {Astropy Collaboration} {et~al.}(2013){Astropy Collaboration},
  {Robitaille}, {Tollerud}, {Greenfield}, {Droettboom}, {Bray}, {Aldcroft},
  {Davis}, {Ginsburg}, {Price-Whelan}, {Kerzendorf}, {Conley}, {Crighton},
  {Barbary}, {Muna}, {Ferguson}, {Grollier}, {Parikh}, {Nair}, {Unther},
  {Deil}, {Woillez}, {Conseil}, {Kramer}, {Turner}, {Singer}, {Fox}, {Weaver},
  {Zabalza}, {Edwards}, {Azalee Bostroem}, {Burke}, {Casey}, {Crawford},
  {Dencheva}, {Ely}, {Jenness}, {Labrie}, {Lim}, {Pierfederici}, {Pontzen},
  {Ptak}, {Refsdal}, {Servillat}, \& {Streicher}}]{astropy:2013}
{Astropy Collaboration}, {Robitaille}, T.~P., {Tollerud}, E.~J., {et~al.} 2013,
  \bibinfo{title}{{Astropy: A community Python package for astronomy},} \aap,
  558, A33, \dodoi{10.1051/0004-6361/201322068}

\bibitem[{ {Astropy Collaboration} {et~al.}(2018){Astropy Collaboration},
  {Price-Whelan}, {Sip{\H{o}}cz}, {G{\"u}nther}, {Lim}, {Crawford}, {Conseil},
  {Shupe}, {Craig}, {Dencheva}, {Ginsburg}, {Vand erPlas}, {Bradley},
  {P{\'e}rez-Su{\'a}rez}, {de Val-Borro}, {Aldcroft}, {Cruz}, {Robitaille},
  {Tollerud}, {Ardelean}, {Babej}, {Bach}, {Bachetti}, {Bakanov}, {Bamford},
  {Barentsen}, {Barmby}, {Baumbach}, {Berry}, {Biscani}, {Boquien}, {Bostroem},
  {Bouma}, {Brammer}, {Bray}, {Breytenbach}, {Buddelmeijer}, {Burke},
  {Calderone}, {Cano Rodr{\'\i}guez}, {Cara}, {Cardoso}, {Cheedella}, {Copin},
  {Corrales}, {Crichton}, {D'Avella}, {Deil}, {Depagne}, {Dietrich}, {Donath},
  {Droettboom}, {Earl}, {Erben}, {Fabbro}, {Ferreira}, {Finethy}, {Fox},
  {Garrison}, {Gibbons}, {Goldstein}, {Gommers}, {Greco}, {Greenfield},
  {Groener}, {Grollier}, {Hagen}, {Hirst}, {Homeier}, {Horton}, {Hosseinzadeh},
  {Hu}, {Hunkeler}, {Ivezi{\'c}}, {Jain}, {Jenness}, {Kanarek}, {Kendrew},
  {Kern}, {Kerzendorf}, {Khvalko}, {King}, {Kirkby}, {Kulkarni}, {Kumar},
  {Lee}, {Lenz}, {Littlefair}, {Ma}, {Macleod}, {Mastropietro}, {McCully},
  {Montagnac}, {Morris}, {Mueller}, {Mumford}, {Muna}, {Murphy}, {Nelson},
  {Nguyen}, {Ninan}, {N{\"o}the}, {Ogaz}, {Oh}, {Parejko}, {Parley}, {Pascual},
  {Patil}, {Patil}, {Plunkett}, {Prochaska}, {Rastogi}, {Reddy Janga},
  {Sabater}, {Sakurikar}, {Seifert}, {Sherbert}, {Sherwood-Taylor}, {Shih},
  {Sick}, {Silbiger}, {Singanamalla}, {Singer}, {Sladen}, {Sooley},
  {Sornarajah}, {Streicher}, {Teuben}, {Thomas}, {Tremblay}, {Turner},
  {Terr{\'o}n}, {van Kerkwijk}, {de la Vega}, {Watkins}, {Weaver}, {Whitmore},
  {Woillez}, {Zabalza}, \& {Astropy Contributors}}]{astropy:2018}
{Astropy Collaboration}, {Price-Whelan}, A.~M., {Sip{\H{o}}cz}, B.~M., {et~al.}
  2018, \bibinfo{title}{{The Astropy Project: Building an Open-science Project
  and Status of the v2.0 Core Package},} \aj, 156, 123,
  \dodoi{10.3847/1538-3881/aabc4f}

\bibitem[{ {Astropy Collaboration} {et~al.}(2022){Astropy Collaboration},
  {Price-Whelan}, {Lim}, {Earl}, {Starkman}, {Bradley}, {Shupe}, {Patil},
  {Corrales}, {Brasseur}, {N{"o}the}, {Donath}, {Tollerud}, {Morris},
  {Ginsburg}, {Vaher}, {Weaver}, {Tocknell}, {Jamieson}, {van Kerkwijk},
  {Robitaille}, {Merry}, {Bachetti}, {G{"u}nther}, {Aldcroft},
  {Alvarado-Montes}, {Archibald}, {B{'o}di}, {Bapat}, {Barentsen}, {Baz{'a}n},
  {Biswas}, {Boquien}, {Burke}, {Cara}, {Cara}, {Conroy}, {Conseil}, {Craig},
  {Cross}, {Cruz}, {D'Eugenio}, {Dencheva}, {Devillepoix}, {Dietrich},
  {Eigenbrot}, {Erben}, {Ferreira}, {Foreman-Mackey}, {Fox}, {Freij}, {Garg},
  {Geda}, {Glattly}, {Gondhalekar}, {Gordon}, {Grant}, {Greenfield}, {Groener},
  {Guest}, {Gurovich}, {Handberg}, {Hart}, {Hatfield-Dodds}, {Homeier},
  {Hosseinzadeh}, {Jenness}, {Jones}, {Joseph}, {Kalmbach}, {Karamehmetoglu},
  {Ka{l}uszy{'n}ski}, {Kelley}, {Kern}, {Kerzendorf}, {Koch}, {Kulumani},
  {Lee}, {Ly}, {Ma}, {MacBride}, {Maljaars}, {Muna}, {Murphy}, {Norman},
  {O'Steen}, {Oman}, {Pacifici}, {Pascual}, {Pascual-Granado}, {Patil},
  {Perren}, {Pickering}, {Rastogi}, {Roulston}, {Ryan}, {Rykoff}, {Sabater},
  {Sakurikar}, {Salgado}, {Sanghi}, {Saunders}, {Savchenko}, {Schwardt},
  {Seifert-Eckert}, {Shih}, {Jain}, {Shukla}, {Sick}, {Simpson},
  {Singanamalla}, {Singer}, {Singhal}, {Sinha}, {Sip{H{o}}cz}, {Spitler},
  {Stansby}, {Streicher}, {{{S}}umak}, {Swinbank}, {Taranu}, {Tewary},
  {Tremblay}, {Val-Borro}, {Van Kooten}, {Vasovi{'c}}, {Verma}, {de Miranda
  Cardoso}, {Williams}, {Wilson}, {Winkel}, {Wood-Vasey}, {Xue}, {Yoachim},
  {Zhang}, {Zonca}, \& {Astropy Project Contributors}}]{astropy:2022}
{Astropy Collaboration}, {Price-Whelan}, A.~M., {Lim}, P.~L., {et~al.} 2022,
  \bibinfo{title}{{The Astropy Project: Sustaining and Growing a
  Community-oriented Open-source Project and the Latest Major Release (v5.0) of
  the Core Package},} \apj, 935, 167, \dodoi{10.3847/1538-4357/ac7c74}

\bibitem[{R. {Baer-Way} {et~al.}(2025){Baer-Way}, {Chandra}, {Modjaz}, {Kumar},
  {Pellegrino}, {Chevalier}, {Crawford}, {Sarangi}, {Smith}, {Maeda}, {Nayana},
  {Filippenko}, {Andrews}, {Arcavi}, {Bostroem}, {Brink}, {Dong}, {Dwarkadas},
  {Farah}, {Howell}, {Hiramatsu}, {Hosseinzadeh}, {McCully}, {Meza}, {Newsome},
  {Padilla Gonzalez}, {Pearson}, {Sand}, {Shrestha}, {Terreran}, {Valenti},
  {Wyatt}, {Yang}, \& {Zheng}}]{Baer25}
{Baer-Way}, R., {Chandra}, P., {Modjaz}, M., {et~al.} 2025, \bibinfo{title}{{A
  Multiwavelength Autopsy of the Interacting Type IIn Supernova 2020ywx:
  Tracing Its Progenitor Mass-loss History for 100 Yr Before Death},} \apj,
  983, 101, \dodoi{10.3847/1538-4357/adc00a}

\bibitem[{S. Bose {et~al.}(2021)Bose, Dong, Kochanek, Stritzinger, Ashall,
  Benetti, Falco, Filippenko, Pastorello, Prieto, Somero, Sukhbold, Zhang,
  Auchettl, Brink, Brown, Chen, Fiore, Grupe, Holoien, Lundqvist, Mattila,
  Mutel, Pooley, Post, Reddy, Reynolds, Shappee, Stanek, Thompson, Villanueva,
  \& Zheng}]{10.1093/mnras/stab629}
Bose, S., Dong, S., Kochanek, C.~S., {et~al.} 2021,
  \bibinfo{title}{ASASSN-18am/SN 2018gk: an overluminous Type IIb supernova
  from a massive progenitor,} Monthly Notices of the Royal Astronomical
  Society, 503, 3472, \dodoi{10.1093/mnras/stab629}

\bibitem[{K.~A. {Bostroem} {et~al.}(2025){Bostroem}, {Valenti}, {Sand},
  {Pearson}, {Shrestha}, {Andrews}, {Dessart}, {Jacobson-Galan}, {Hsu}, {Ravi},
  {Andrews}, {Christy}, {Dong}, {Franz}, {Farah}, {Filippenko}, {Gill},
  {Hoang}, {Hosseinzadeh}, {Howell}, {Janzen}, {Jencson}, {Jha}, {Kwok},
  {Lundquist}, {Martas}, {McCully}, {Mehta}, {Newsome}, {Padilla-Gonzalez},
  {Meza Retamal}, {Smith}, {Subrayan}, \& {Terreran}}]{Bostroem25}
{Bostroem}, K.~A., {Valenti}, S., {Sand}, D.~J., {et~al.} 2025,
  \bibinfo{title}{{Late-time Hubble Space Telescope Ultraviolet Spectra of SN
  2023ixf and SN 2024ggi Show Ongoing Interaction with Circumstellar
  Material},} arXiv e-prints, arXiv:2508.11756,
  \dodoi{10.48550/arXiv.2508.11756}

\bibitem[{D. Brethauer {et~al.}(2022)Brethauer, Margutti, Milisavljevic,
  Bietenholz, Chornock, Coppejans, Colle, Hajela, Terreran, Vargas, DeMarchi,
  Harris, Jacobson-Galán, Kamble, Patnaude, \& Stroh}]{Brethauer_2022}
Brethauer, D., Margutti, R., Milisavljevic, D., {et~al.} 2022,
  \bibinfo{title}{Seven Years of Coordinated Chandra–NuSTAR Observations of
  SN 2014C Unfold the Extreme Mass-loss History of Its Stellar Progenitor,} The
  Astrophysical Journal, 939, 105, \dodoi{10.3847/1538-4357/ac8b14}

\bibitem[{J. {Buchner} {et~al.}(2014){Buchner}, {Georgakakis}, {Nandra}, {Hsu},
  {Rangel}, {Brightman}, {Merloni}, {Salvato}, {Donley}, \&
  {Kocevski}}]{2014A&A...564A.125B}
{Buchner}, J., {Georgakakis}, A., {Nandra}, K., {et~al.} 2014,
  \bibinfo{title}{{X-ray spectral modelling of the AGN obscuring region in the
  CDFS: Bayesian model selection and catalogue},} \aap, 564, A125,
  \dodoi{10.1051/0004-6361/201322971}

\bibitem[{D.~N. {Burrows} {et~al.}(2005){Burrows}, {Hill}, {Nousek}, {Kennea},
  {Wells}, {Osborne}, {Abbey}, {Beardmore}, {Mukerjee}, {Short}, {Chincarini},
  {Campana}, {Citterio}, {Moretti}, {Pagani}, {Tagliaferri}, {Giommi},
  {Capalbi}, {Tamburelli}, {Angelini}, {Cusumano}, {Br{\"a}uninger}, {Burkert},
  \& {Hartner}}]{2005SSRv..120..165B}
{Burrows}, D.~N., {Hill}, J.~E., {Nousek}, J.~A., {et~al.} 2005,
  \bibinfo{title}{{The Swift X-Ray Telescope},} \ssr, 120, 165,
  \dodoi{10.1007/s11214-005-5097-2}

\bibitem[{S. Chakraborti {et~al.}(2013)Chakraborti, Ray, Smith, Ryder, Yadav,
  Sutaria, Dwarkadas, Chandra, Pooley, \& Roy}]{Chakraborti_2013}
Chakraborti, S., Ray, A., Smith, R., {et~al.} 2013, \bibinfo{title}{THE
  PROGENITOR OF SN 2011ja: CLUES FROM CIRCUMSTELLAR INTERACTION,} The
  Astrophysical Journal, 774, 30, \dodoi{10.1088/0004-637X/774/1/30}

\bibitem[{S. Chakraborti {et~al.}(2016)Chakraborti, Ray, Smith, Margutti,
  Pooley, Bose, Sutaria, Chandra, Dwarkadas, Ryder, \&
  Maeda}]{Chakraborti_2016}
Chakraborti, S., Ray, A., Smith, R., {et~al.} 2016, \bibinfo{title}{PROBING
  FINAL STAGES OF STELLAR EVOLUTION WITH X-RAY OBSERVATIONS OF SN 2013ej,} The
  Astrophysical Journal, 817, 22, \dodoi{10.3847/0004-637X/817/1/22}

\bibitem[{P. Chandra {et~al.}(2022)Chandra, Chevalier, James, \&
  Fox}]{10.1093/mnras/stac2915}
Chandra, P., Chevalier, R.~A., James, N. J.~H., \& Fox, O.~D. 2022,
  \bibinfo{title}{The luminous type IIn supernova SN 2017hcc: Infrared bright,
  X-ray, and radio faint,} Monthly Notices of the Royal Astronomical Society,
  517, 4151, \dodoi{10.1093/mnras/stac2915}

\bibitem[{P. {Chandra} {et~al.}(2024{\natexlab{a}}){Chandra}, {Chevalier},
  {Maeda}, {Ray}, \& {Nayana}}]{2024ApJ...963L...4C}
{Chandra}, P., {Chevalier}, R.~A., {Maeda}, K., {Ray}, A.~K., \& {Nayana},
  A.~J. 2024{\natexlab{a}}, \bibinfo{title}{{Chandra's Insights into SN
  2023ixf},} \apjl, 963, L4, \dodoi{10.3847/2041-8213/ad275d}

\bibitem[{P. {Chandra} {et~al.}(2024{\natexlab{b}}){Chandra}, {Maeda},
  {Nayana}, {Chevalier}, {Ryder}, {Ho}, \& {Ray}}]{2024ATel16612....1C}
{Chandra}, P., {Maeda}, K., {Nayana}, A.~J., {et~al.} 2024{\natexlab{b}},
  \bibinfo{title}{{Centimeter-wavelength upper limits on SN 2024ggi with the
  JVLA and the uGMRT},} The Astronomer's Telegram, 16612, 1

\bibitem[{T.~W. {Chen} {et~al.}(2024){Chen}, {K}, {Yang}, {Srivastav},
  {Smartt}, {Hou}, {Lin}, {Brennan}, {Lee}, {Lai}, {Aryan}, {Hsiao}, {Pan},
  {Ngeow}, {Lin}, {Guo}, {Stevance}, {Gillanders}, {Smith}, {Fulton}, {Moore},
  {Angus}, \& {Aamer}}]{2024TNSAN.102....1C}
{Chen}, T.~W., {K}, A.~S., {Yang}, S., {et~al.} 2024, \bibinfo{title}{{Kinder
  follow-up observations of AT 2024ggi (ATLAS24fsk)},} Transient Name Server
  AstroNote, 102, 1

\bibitem[{T.~W. {Chen} {et~al.}(2025){Chen}, {Yang}, {Srivastav}, {Moriya},
  {Smartt}, {Rest}, {Rest}, {Lin}, {Miao}, {Cheng}, {Aryan}, {Cheng}, {Fraser},
  {Huang}, {Lee}, {Lai}, {Liu}, {Sankar. K}, {Smith}, {Stevance}, {Wang},
  {Anderson}, {Angus}, {de Boer}, {Chambers}, {Duan}, {Erasmus}, {Fulton},
  {Gao}, {Herman}, {Hou}, {Hsiao}, {Huber}, {Lin}, {Lin}, {Magnier}, {Man},
  {Moore}, {Ngeow}, {Nicholl}, {Ou}, {Pignata}, {Shiau}, {Sommer}, {Tonry},
  {Wang}, {Wainscoat}, {Young}, {Yeh}, \& {Zhang}}]{2025ApJ...983...86C}
{Chen}, T.~W., {Yang}, S., {Srivastav}, S., {et~al.} 2025,
  \bibinfo{title}{{Discovery and Extensive Follow-up of SN 2024ggi, a Nearby
  Type IIP Supernova in NGC 3621},} \apj, 983, 86,
  \dodoi{10.3847/1538-4357/adb428}

\bibitem[{R.~A.
  {Chevalier}(1982{\natexlab{a}}){Chevalier}}]{1982ApJ...259..302C}
{Chevalier}, R.~A. 1982{\natexlab{a}}, \bibinfo{title}{{The radio and X-ray
  emission from type II supernovae.},} \apj, 259, 302, \dodoi{10.1086/160167}

\bibitem[{R.~A.
  {Chevalier}(1982{\natexlab{b}}){Chevalier}}]{1982ApJ...258..790C}
{Chevalier}, R.~A. 1982{\natexlab{b}}, \bibinfo{title}{{Self-similar solutions
  for the interaction of stellar ejecta with an external medium.},} \apj, 258,
  790, \dodoi{10.1086/160126}

\bibitem[{R.~A. Chevalier \& C. Fransson(2017)Chevalier \&
  Fransson}]{Chevalier2017}
Chevalier, R.~A., \& Fransson, C. 2017, Thermal and Non-thermal Emission from
  Circumstellar Interaction, ed. A.~W. Alsabti \& P.~Murdin (Cham: Springer
  International Publishing), 875--937, \dodoi{10.1007/978-3-319-21846-5_34}

\bibitem[{L. {Dessart} {et~al.}(2025){Dessart}, {Kotak}, {Jacobson-Galan},
  {Das}, {Fremling}, {Kasliwal}, {Qin}, \& {Rose}}]{2025arXiv250705803D}
{Dessart}, L., {Kotak}, R., {Jacobson-Galan}, W., {et~al.} 2025,
  \bibinfo{title}{{An optical to infrared study of type II SN2024ggi at nebular
  times},} arXiv e-prints, arXiv:2507.05803, \dodoi{10.48550/arXiv.2507.05803}

\bibitem[{V.~V. Dwarkadas(2025)Dwarkadas}]{universe11050161}
Dwarkadas, V.~V. 2025, \bibinfo{title}{On the X-Ray Emission from Supernovae,
  and Implications for the Mass-Loss Rates of Their Progenitor Stars,}
  Universe, 11, \dodoi{10.3390/universe11050161}

\bibitem[{K. {Ertini} {et~al.}(2025){Ertini}, {Regna}, {Ferrari}, {Bersten},
  {Folatelli}, {Mendez Llorca}, {Fern{\'a}ndez-Laj{\'u}s}, {Ferrero},
  {Hueichap{\'a}n D{\'\i}az}, {Cartier}, {Rom{\'a}n Aguilar}, {Putkuri},
  {Piccirilli}, {Cellone}, {Moreno}, {Orellana}, {Prieto}, {Gerlach}, {Acosta},
  {Ritacco}, {Schujman}, \& {Vald{\'e}z}}]{2025A&A...699A..60E}
{Ertini}, K., {Regna}, T.~A., {Ferrari}, L., {et~al.} 2025, \bibinfo{title}{{SN
  2024ggi: Another year, another striking Type II supernova},} \aap, 699, A60,
  \dodoi{10.1051/0004-6361/202554333}

\bibitem[{P.~A. {Evans} {et~al.}(2009){Evans}, {Beardmore}, {Page}, {Osborne},
  {O'Brien}, {Willingale}, {Starling}, {Burrows}, {Godet}, {Vetere}, {Racusin},
  {Goad}, {Wiersema}, {Angelini}, {Capalbi}, {Chincarini}, {Gehrels}, {Kennea},
  {Margutti}, {Morris}, {Mountford}, {Pagani}, {Perri}, {Romano}, \&
  {Tanvir}}]{2009MNRAS.397.1177E}
{Evans}, P.~A., {Beardmore}, A.~P., {Page}, K.~L., {et~al.} 2009,
  \bibinfo{title}{{Methods and results of an automatic analysis of a complete
  sample of Swift-XRT observations of GRBs},} \mnras, 397, 1177,
  \dodoi{10.1111/j.1365-2966.2009.14913.x}

\bibitem[{L. {Ferrari} {et~al.}(2025){Ferrari}, {Folatelli}, {Ertini},
  {Kuncarayakti}, {Regna}, {Bersten}, {Ashall}, {Baron}, {Burns}, {Galbany},
  {Hoogendam}, {Maeda}, {Medler}, {Morrell}, {Shappee}, {Stritzinger}, \&
  {Xiao}}]{Ferrari25}
{Ferrari}, L., {Folatelli}, G., {Ertini}, K., {et~al.} 2025,
  \bibinfo{title}{{The nebular phase of SN 2024ggi: A low-mass progenitor with
  no signs of interaction},} \aap, 703, A12,
  \dodoi{10.1051/0004-6361/202556652}

\bibitem[{D. Foreman-Mackey(2016)Foreman-Mackey}]{corner}
Foreman-Mackey, D. 2016, \bibinfo{title}{corner.py: Scatterplot matrices in
  Python,} The Journal of Open Source Software, 1, 24,
  \dodoi{10.21105/joss.00024}

\bibitem[{D. {Foreman-Mackey} {et~al.}(2013){Foreman-Mackey}, {Hogg}, {Lang},
  \& {Goodman}}]{2013PASP..125..306F}
{Foreman-Mackey}, D., {Hogg}, D.~W., {Lang}, D., \& {Goodman}, J. 2013,
  \bibinfo{title}{{emcee: The MCMC Hammer},} \pasp, 125, 306,
  \dodoi{10.1086/670067}

\bibitem[{O.~D. Fox {et~al.}(2014)Fox, Silverman, Filippenko, Mauerhan, Becker,
  Borish, Cenko, Clubb, Graham, Hsiao, Kelly, Lee, Marion, Milisavljevic,
  Parrent, Shivvers, Skrutskie, Smith, Wilson, \&
  Zheng}]{10.1093/mnras/stu2435}
Fox, O.~D., Silverman, J.~M., Filippenko, A.~V., {et~al.} 2014,
  \bibinfo{title}{On the nature of Type IIn/Ia–CSM supernovae: optical and
  near-infrared spectra of SN 2012ca and SN 2013dn,} Monthly Notices of the
  Royal Astronomical Society, 447, 772, \dodoi{10.1093/mnras/stu2435}

\bibitem[{C. Fransson {et~al.}(2014)Fransson, Ergon, Challis, Chevalier,
  France, Kirshner, Marion, Milisavljevic, Smith, Bufano, Friedman, Kangas,
  Larsson, Mattila, Benetti, Chornock, Czekala, Soderberg, \&
  Sollerman}]{Fransson_2014}
Fransson, C., Ergon, M., Challis, P.~J., {et~al.} 2014,
  \bibinfo{title}{HIGH-DENSITY CIRCUMSTELLAR INTERACTION IN THE LUMINOUS TYPE
  IIn SN 2010jl: THE FIRST 1100 DAYS,} The Astrophysical Journal, 797, 118,
  \dodoi{10.1088/0004-637X/797/2/118}

\bibitem[{A. {Fruscione} {et~al.}(2006){Fruscione}, {McDowell}, {Allen},
  {Brickhouse}, {Burke}, {Davis}, {Durham}, {Elvis}, {Galle}, {Harris},
  {Huenemoerder}, {Houck}, {Ishibashi}, {Karovska}, {Nicastro}, {Noble},
  {Nowak}, {Primini}, {Siemiginowska}, {Smith}, \&
  {Wise}}]{2006SPIE.6270E..1VF}
{Fruscione}, A., {McDowell}, J.~C., {Allen}, G.~E., {et~al.} 2006, in Society
  of Photo-Optical Instrumentation Engineers (SPIE) Conference Series, Vol.
  6270, Observatory Operations: Strategies, Processes, and Systems, ed. D.~R.
  {Silva} \& R.~E. {Doxsey}, 62701V, \dodoi{10.1117/12.671760}

\bibitem[{J. {Fuller} \& D. {Tsuna}(2024){Fuller} \&
  {Tsuna}}]{2024OJAp....7E..47F}
{Fuller}, J., \& {Tsuna}, D. 2024, \bibinfo{title}{{Boil-off of red
  supergiants: mass loss and type II-P supernovae},} The Open Journal of
  Astrophysics, 7, 47, \dodoi{10.33232/001c.120130}

\bibitem[{N. {Gehrels} {et~al.}(2004){Gehrels}, {Chincarini}, {Giommi},
  {Mason}, {Nousek}, {Wells}, {White}, {Barthelmy}, {Burrows}, {Cominsky},
  {Hurley}, {Marshall}, {M{\'e}sz{\'a}ros}, {Roming}, {Angelini}, {Barbier},
  {Belloni}, {Campana}, {Caraveo}, {Chester}, {Citterio}, {Cline}, {Cropper},
  {Cummings}, {Dean}, {Feigelson}, {Fenimore}, {Frail}, {Fruchter}, {Garmire},
  {Gendreau}, {Ghisellini}, {Greiner}, {Hill}, {Hunsberger}, {Krimm},
  {Kulkarni}, {Kumar}, {Lebrun}, {Lloyd-Ronning}, {Markwardt}, {Mattson},
  {Mushotzky}, {Norris}, {Osborne}, {Paczynski}, {Palmer}, {Park}, {Parsons},
  {Paul}, {Rees}, {Reynolds}, {Rhoads}, {Sasseen}, {Schaefer}, {Short},
  {Smale}, {Smith}, {Stella}, {Tagliaferri}, {Takahashi}, {Tashiro},
  {Townsley}, {Tueller}, {Turner}, {Vietri}, {Voges}, {Ward}, {Willingale},
  {Zerbi}, \& {Zhang}}]{2004ApJ...611.1005G}
{Gehrels}, N., {Chincarini}, G., {Giommi}, P., {et~al.} 2004,
  \bibinfo{title}{{The Swift Gamma-Ray Burst Mission},} \apj, 611, 1005,
  \dodoi{10.1086/422091}

\bibitem[{C. {Gordon} \& K. {Arnaud}(2021){Gordon} \&
  {Arnaud}}]{2021ascl.soft01014G}
{Gordon}, C., \& {Arnaud}, K. 2021, \bibinfo{title}{{PyXspec: Python interface
  to XSPEC spectral-fitting program},}, Astrophysics Source Code Library,
  record ascl:2101.014

\bibitem[{B.~W. {Grefenstette} {et~al.}(2023){Grefenstette}, {Brightman},
  {Earnshaw}, {Harrison}, \& {Margutti}}]{2023ApJ...952L...3G}
{Grefenstette}, B.~W., {Brightman}, M., {Earnshaw}, H.~P., {Harrison}, F.~A.,
  \& {Margutti}, R. 2023, \bibinfo{title}{{Early Hard X-Rays from the Nearby
  Core-collapse Supernova SN 2023ixf},} \apjl, 952, L3,
  \dodoi{10.3847/2041-8213/acdf4e}

\bibitem[{C.~R. Harris {et~al.}(2020)Harris, Millman, van~der Walt, Gommers,
  Virtanen, Cournapeau, Wieser, Taylor, Berg, Smith, Kern, Picus, Hoyer, van
  Kerkwijk, Brett, Haldane, del R{\'{i}}o, Wiebe, Peterson,
  G{\'{e}}rard-Marchant, Sheppard, Reddy, Weckesser, Abbasi, Gohlke, \&
  Oliphant}]{harris2020array}
Harris, C.~R., Millman, K.~J., van~der Walt, S.~J., {et~al.} 2020,
  \bibinfo{title}{Array programming with {NumPy},} Nature, 585, 357,
  \dodoi{10.1038/s41586-020-2649-2}

\bibitem[{F.~A. Harrison {et~al.}(2013)Harrison, Craig, Christensen, Hailey,
  Zhang, Boggs, Stern, Cook, Forster, Giommi, Grefenstette, Kim, Kitaguchi,
  Koglin, Madsen, Mao, Miyasaka, Mori, Perri, Pivovaroff, Puccetti, Rana,
  Westergaard, Willis, Zoglauer, An, Bachetti, Barrière, Bellm, Bhalerao,
  Brejnholt, Fuerst, Liebe, Markwardt, Nynka, Vogel, Walton, Wik, Alexander,
  Cominsky, Hornschemeier, Hornstrup, Kaspi, Madejski, Matt, Molendi, Smith,
  Tomsick, Ajello, Ballantyne, Baloković, Barret, Bauer, Blandford, Brandt,
  Brenneman, Chiang, Chakrabarty, Chenevez, Comastri, Dufour, Elvis, Fabian,
  Farrah, Fryer, Gotthelf, Grindlay, Helfand, Krivonos, Meier, Miller,
  Natalucci, Ogle, Ofek, Ptak, Reynolds, Rigby, Tagliaferri, Thorsett,
  Treister, \& Urry}]{Harrison_2013}
Harrison, F.~A., Craig, W.~W., Christensen, F.~E., {et~al.} 2013,
  \bibinfo{title}{THE NUCLEAR SPECTROSCOPIC TELESCOPE ARRAY (NuSTAR)
  HIGH-ENERGY X-RAY MISSION,} The Astrophysical Journal, 770, 103,
  \dodoi{10.1088/0004-637X/770/2/103}

\bibitem[{ {HI4PI Collaboration} {et~al.}(2016){HI4PI Collaboration}, {Ben
  Bekhti}, {Fl{\"o}er}, {Keller}, {Kerp}, {Lenz}, {Winkel}, {Bailin},
  {Calabretta}, {Dedes}, {Ford}, {Gibson}, {Haud}, {Janowiecki}, {Kalberla},
  {Lockman}, {McClure-Griffiths}, {Murphy}, {Nakanishi}, {Pisano}, \&
  {Staveley-Smith}}]{2016A&A...594A.116H}
{HI4PI Collaboration}, {Ben Bekhti}, N., {Fl{\"o}er}, L., {et~al.} 2016,
  \bibinfo{title}{{HI4PI: A full-sky H I survey based on EBHIS and GASS},}
  \aap, 594, A116, \dodoi{10.1051/0004-6361/201629178}

\bibitem[{W. {Hoogendam} {et~al.}(2024){Hoogendam}, {Auchettl}, {Tucker},
  {Ashall}, {Shappee}, \& {Huber}}]{2024TNSAN.103....1H}
{Hoogendam}, W., {Auchettl}, K., {Tucker}, M., {et~al.} 2024,
  \bibinfo{title}{{Early Spectrum of SN 2024ggi from SCAT shows a Type II SN
  with Flash Ionized Features},} Transient Name Server AstroNote, 103, 1

\bibitem[{E. {Hueichap{\'a}n} {et~al.}(2025){Hueichap{\'a}n}, {Cartier},
  {Prieto}, {Contreras}, {Cikota}, {Pessi}, {Bauer}, \& {Pignata}}]{Hueichap25}
{Hueichap{\'a}n}, E., {Cartier}, R., {Prieto}, J.~L., {et~al.} 2025,
  \bibinfo{title}{{Optical and near-infrared nebular-phase spectroscopy of SN
  2024ggi: constraints on the structure of the inner ejecta, progenitor mass,
  and dust},} arXiv e-prints, arXiv:2508.02656,
  \dodoi{10.48550/arXiv.2508.02656}

\bibitem[{S. {Immler} \& P.~J. {Brown}(2012){Immler} \&
  {Brown}}]{2012ATel.3995....1I}
{Immler}, S., \& {Brown}, P.~J. 2012, \bibinfo{title}{{Swift XRT Detection of
  Supernova 2012aw in X-Rays},} The Astronomer's Telegram, 3995, 1

\bibitem[{S. Immler {et~al.}(2007)Immler, Brown, Milne, Dessart, Mazzali,
  Landsman, Gehrels, Petre, Burrows, Nousek, Chevalier, Williams, Koss,
  Stockdale, Kelley, Weiler, Holland, Pian, Roming, Pooley, Nomoto, Greiner,
  Campana, \& Soderberg}]{Immler_2007}
Immler, S., Brown, P.~J., Milne, P., {et~al.} 2007, \bibinfo{title}{X-Ray, UV,
  and Optical Observations of Supernova 2006bp with Swift: Detection of Early
  X-Ray Emission,} The Astrophysical Journal, 664, 435, \dodoi{10.1086/518466}

\bibitem[{I. {Irani} {et~al.}(2024){Irani}, {Morag}, {Gal-Yam}, {Waxman},
  {Schulze}, {Sollerman}, {Hinds}, {Perley}, {Chen}, {Strotjohann}, {Yaron},
  {Zimmerman}, {Bruch}, {Ofek}, {Soumagnac}, {Yang}, {Groom}, {Masci},
  {Aubert}, {Riddle}, {Bellm}, \& {Hale}}]{2024ApJ...970...96I}
{Irani}, I., {Morag}, J., {Gal-Yam}, A., {et~al.} 2024, \bibinfo{title}{{The
  Early Ultraviolet Light Curves of Type II Supernovae and the Radii of Their
  Progenitor Stars},} \apj, 970, 96, \dodoi{10.3847/1538-4357/ad3de8}

\bibitem[{K. {Itagaki}(2023){Itagaki}}]{2023TNSTR1158....1I}
{Itagaki}, K. 2023, \bibinfo{title}{{Transient Discovery Report for
  2023-05-19},} Transient Name Server Discovery Report, 2023-1158, 1

\bibitem[{W.
  {Jacobson-Gal{\'a}n}(2025){Jacobson-Gal{\'a}n}}]{2025Univ...11..231J}
{Jacobson-Gal{\'a}n}, W. 2025, \bibinfo{title}{{SN 2023ixf: The Closest
  Supernova of the Decade},} Universe, 11, 231,
  \dodoi{10.3390/universe11070231}

\bibitem[{W.~V. {Jacobson-Gal{\'a}n}
  {et~al.}(2025{\natexlab{a}}){Jacobson-Gal{\'a}n}, {Dessart}, {Kilpatrick},
  {Patel}, {Auchettl}, {Tinyanont}, {Margutti}, {Dwarkadas}, {Bostroem},
  {Chornock}, {Foley}, {Abunemeh}, {Ahumada}, {Arunachalam},
  {Bustamante-Rosell}, {Coulter}, {Gall}, {Gao}, {Guo}, {Hjorth}, {Kaewmookda},
  {Kasliwal}, {Kaur}, {Larison}, {LeBaron}, {Miao}, {Narayan}, {Pan}, {Park},
  {Patra}, {Qin}, {Ransome}, {Rest}, {Rho}, {Rose}, {Sears}, {Swift},
  {Taggart}, {Villar}, {Wang}, {Zenati}, \& {Zhou}}]{WJG25c}
{Jacobson-Gal{\'a}n}, W.~V., {Dessart}, L., {Kilpatrick}, C.~D., {et~al.}
  2025{\natexlab{a}}, \bibinfo{title}{{A Panchromatic View of Late-time Shock
  Power in the Type II Supernova 2023ixf},} arXiv e-prints, arXiv:2508.11747,
  \dodoi{10.48550/arXiv.2508.11747}

\bibitem[{W.~V. {Jacobson-Gal{\'a}n}
  {et~al.}(2025{\natexlab{b}}){Jacobson-Gal{\'a}n}, {Dessart}, {Davis},
  {Bostroem}, {Kilpatrick}, {Margutti}, {Filippenko}, {Foley}, {Chornock},
  {Terreran}, {Hiramatsu}, {Newsome}, {Padilla Gonzalez}, {Pellegrino},
  {Howell}, {Anderson}, {Angus}, {Auchettl}, {Brink}, {Cartier}, {Coulter}, {de
  Boer}, {Drout}, {Earl}, {Ertini}, {Farah}, {Farias}, {Gall}, {Gao},
  {Gerlach}, {Guo}, {Haynie}, {Hosseinzadeh}, {Ibik}, {Jha}, {Jones},
  {Langeroodi}, {LeBaron}, {Magnier}, {Piro}, {Raimundo}, {Rest}, {Rest},
  {Michael Rich}, {Rojas-Bravo}, {Sears}, {Taggart}, {Villar}, {Wainscoat},
  {Wang}, {Wasserman}, {Yan}, {Yang}, {Zhang}, \& {Zheng}}]{WJG25b}
{Jacobson-Gal{\'a}n}, W.~V., {Dessart}, L., {Davis}, K.~W., {et~al.}
  2025{\natexlab{b}}, \bibinfo{title}{{Final Moments. III. Explosion Properties
  and Progenitor Constraints of CSM-interacting Type II Supernovae},} \apj,
  992, 100, \dodoi{10.3847/1538-4357/adfa23}

\bibitem[{W.~V. Jacobson-Galán {et~al.}(2024{\natexlab{a}})Jacobson-Galán,
  Davis, Kilpatrick, Dessart, Margutti, Chornock, Foley, Arunachalam, Auchettl,
  Bom, Cartier, Coulter, Dimitriadis, Dickinson, Drout, Gagliano, Gall,
  Garretson, Izzo, Jones, LeBaron, Miao, Milisavljevic, Pan, Rest, Rojas-Bravo,
  Santos, Sears, Subrayan, Taggart, \& Tinyanont}]{Jacobson-Galan2024}
Jacobson-Galán, W.~V., Davis, K.~W., Kilpatrick, C.~D., {et~al.}
  2024{\natexlab{a}}, \bibinfo{title}{SN 2024ggi in NGC 3621: Rising Ionization
  in a Nearby, Circumstellar-material-interacting Type II Supernova,} The
  Astrophysical Journal, 972, 177, \dodoi{10.3847/1538-4357/ad5c64}

\bibitem[{W.~V. Jacobson-Galán {et~al.}(2024{\natexlab{b}})Jacobson-Galán,
  Dessart, Davis, Kilpatrick, Margutti, Foley, Chornock, Terreran, Hiramatsu,
  Newsome, Padilla~Gonzalez, Pellegrino, Howell, Filippenko, Anderson, Angus,
  Auchettl, Bostroem, Brink, Cartier, Coulter, de~Boer, Drout, Earl, Ertini,
  Farah, Farias, Gall, Gao, Gerlach, Guo, Haynie, Hosseinzadeh, Ibik, Jha,
  Jones, Langeroodi, LeBaron, Magnier, Piro, Raimundo, Rest, Rest, Rich,
  Rojas-Bravo, Sears, Taggart, Villar, Wainscoat, Wang, Wasserman, Yan, Yang,
  Zhang, \& Zheng}]{Jacobson-Galan2024b}
Jacobson-Galán, W.~V., Dessart, L., Davis, K.~W., {et~al.} 2024{\natexlab{b}},
  \bibinfo{title}{Final Moments. II. Observational Properties and Physical
  Modeling of Circumstellar-material-interacting Type II Supernovae,} The
  Astrophysical Journal, 970, 189, \dodoi{10.3847/1538-4357/ad4a2a}

\bibitem[{W.~A. {Joye} \& E. {Mandel}(2003){Joye} \&
  {Mandel}}]{2003ASPC..295..489J}
{Joye}, W.~A., \& {Mandel}, E. 2003, in Astronomical Society of the Pacific
  Conference Series, Vol. 295, Astronomical Data Analysis Software and Systems
  XII, ed. H.~E. {Payne}, R.~I. {Jedrzejewski}, \& R.~N. {Hook}, 489

\bibitem[{S. {Katsuda} {et~al.}(2014){Katsuda}, {Maeda}, {Nozawa}, {Pooley}, \&
  {Immler}}]{2014ApJ...780..184K}
{Katsuda}, S., {Maeda}, K., {Nozawa}, T., {Pooley}, D., \& {Immler}, S. 2014,
  \bibinfo{title}{{SN 2005ip: A Luminous Type IIn Supernova Emerging from a
  Dense Circumstellar Medium as Revealed by X-Ray Observations},} \apj, 780,
  184, \dodoi{10.1088/0004-637X/780/2/184}

\bibitem[{T. {Killestein} {et~al.}(2024){Killestein}, {Ackley}, {Kotak},
  {O'Neill}, {Ramsay}, {Steeghs}, {Pursiainen}, {Dyer}, {Jim{\'e}nez-Ibarra},
  {Lyman}, {Ulaczyk}, {Galloway}, {Dhillon}, {O'Brien}, {Noysena}, {Breton},
  {Nuttall}, {Pall{\'e}}, {Pollacco}, {Kumar}, {O'Neill}, \&
  {Jarvis}}]{2024TNSAN.101....1K}
{Killestein}, T., {Ackley}, K., {Kotak}, R., {et~al.} 2024,
  \bibinfo{title}{{AT2024ggi: confirmation of an infant transient at 7 Mpc and
  improved constraint on explosion epoch},} Transient Name Server AstroNote,
  101, 1

\bibitem[{C.~D. {Kilpatrick} {et~al.}(2025){Kilpatrick}, {Suresh}, {Davis},
  {Drout}, {Foley}, {Gagliano}, {Jacobson-Gal{\'a}n}, {Kaur}, {Taggart}, \&
  {Vazquez}}]{Kilpatrick25}
{Kilpatrick}, C.~D., {Suresh}, A., {Davis}, K.~W., {et~al.} 2025,
  \bibinfo{title}{{The Type II SN 2025pht in NGC 1637: A Red Supergiant with
  Carbon-rich Circumstellar Dust as the First JWST Detection of a Supernova
  Progenitor Star},} \apjl, 992, L10, \dodoi{10.3847/2041-8213/ae04de}

\bibitem[{B. {Kumar} {et~al.}(2024){Kumar}, {Chen}, {Lin}, {Fang}, {Du}, {Er},
  {Liu}, {Zhang}, {Zhang}, {Bao}, {Zou}, {Pan}, {Zhang}, {Chatterjee}, {Liu},
  {Yang}, \& {Liu}}]{2024TNSAN.108....1K}
{Kumar}, B., {Chen}, X., {Lin}, W., {et~al.} 2024, \bibinfo{title}{{Monitoring
  of AT 2024ggi (ATLAS24fsk) with Mephisto facilities},} Transient Name Server
  AstroNote, 108, 1

\bibitem[{K. {Maeda} {et~al.}(2015){Maeda}, {Hattori}, {Milisavljevic},
  {Folatelli}, {Drout}, {Kuncarayakti}, {Margutti}, {Kamble}, {Soderberg},
  {Tanaka}, {Kawabata}, {Kawabata}, {Yamanaka}, {Nomoto}, {Kim}, {Simon},
  {Phillips}, {Parrent}, {Nakaoka}, {Moriya}, {Suzuki}, {Takaki}, {Ishigaki},
  {Sakon}, {Tajitsu}, \& {Iye}}]{2015ApJ...807...35M}
{Maeda}, K., {Hattori}, T., {Milisavljevic}, D., {et~al.} 2015,
  \bibinfo{title}{{Type IIb Supernova 2013df Entering into an Interaction
  Phase: A Link between the Progenitor and the Mass Loss},} \apj, 807, 35,
  \dodoi{10.1088/0004-637X/807/1/35}

\bibitem[{R. {Margutti} \& B. {Grefenstette}(2024){Margutti} \&
  {Grefenstette}}]{2024ATel16587....1M}
{Margutti}, R., \& {Grefenstette}, B. 2024, \bibinfo{title}{{NuSTAR detection
  of SN2024ggi at 2 days post discovery},} The Astronomer's Telegram, 16587, 1

\bibitem[{R. Margutti {et~al.}(2012)Margutti, Zaninoni, Bernardini, Chincarini,
  Pasotti, Guidorzi, Angelini, Burrows, Capalbi, Evans, Gehrels, Kennea,
  Mangano, Moretti, Nousek, Osborne, Page, Perri, Racusin, Romano, Sbarufatti,
  Stafford, \& Stamatikos}]{10.1093/mnras/sts066}
Margutti, R., Zaninoni, E., Bernardini, M.~G., {et~al.} 2012,
  \bibinfo{title}{The prompt-afterglow connection in gamma-ray bursts: a
  comprehensive statistical analysis of Swift X-ray light curves,} Monthly
  Notices of the Royal Astronomical Society, 428, 729,
  \dodoi{10.1093/mnras/sts066}

\bibitem[{R. {Margutti} {et~al.}(2017){Margutti}, {Kamble}, {Milisavljevic},
  {Zapartas}, {de Mink}, {Drout}, {Chornock}, {Risaliti}, {Zauderer},
  {Bietenholz}, {Cantiello}, {Chakraborti}, {Chomiuk}, {Fong}, {Grefenstette},
  {Guidorzi}, {Kirshner}, {Parrent}, {Patnaude}, {Soderberg}, {Gehrels}, \&
  {Harrison}}]{2017ApJ...835..140M}
{Margutti}, R., {Kamble}, A., {Milisavljevic}, D., {et~al.} 2017,
  \bibinfo{title}{{Ejection of the Massive Hydrogen-rich Envelope Timed with
  the Collapse of the Stripped SN 2014C},} \apj, 835, 140,
  \dodoi{10.3847/1538-4357/835/2/140}

\bibitem[{G. {Marti-Devesa} \&  {Fermi-LAT Collaboration}(2024){Marti-Devesa}
  \& {Fermi-LAT Collaboration}}]{2024ATel16601....1M}
{Marti-Devesa}, G., \& {Fermi-LAT Collaboration}. 2024,
  \bibinfo{title}{{Fermi-LAT gamma-ray observations of SN 2024ggi},} The
  Astronomer's Telegram, 16601, 1

\bibitem[{K. Misra {et~al.}(2007)Misra, Pooley, Chandra, Bhattacharya, Ray,
  Sagar, \& Lewin}]{10.1111/j.1365-2966.2007.12258.x}
Misra, K., Pooley, D., Chandra, P., {et~al.} 2007, \bibinfo{title}{Type IIP
  supernova SN 2004et: a multiwavelength study in X-ray, optical and radio,}
  Monthly Notices of the Royal Astronomical Society, 381, 280,
  \dodoi{10.1111/j.1365-2966.2007.12258.x}

\bibitem[{ {Nasa High Energy Astrophysics Science Archive Research Center
  (Heasarc)}(2014){Nasa High Energy Astrophysics Science Archive Research
  Center (Heasarc)}}]{2014ascl.soft08004N}
{Nasa High Energy Astrophysics Science Archive Research Center (Heasarc)}.
  2014, \bibinfo{title}{{HEAsoft: Unified Release of FTOOLS and XANADU},},
  Astrophysics Source Code Library, record ascl:1408.004

\bibitem[{A.~J. Nayana {et~al.}(2025)Nayana, Margutti, Wiston, Chornock,
  Campana, Laskar, Murase, Krips, Migliori, Tsuna, Alexander, Chandra,
  Bietenholz, Berger, Chevalier, De~Colle, Dessart, Diesing, Grefenstette,
  Jacobson-Galán, Maeda, Marcote, Matthews, Milisavljevic, Ray, Reguitti, \&
  Polzin}]{AJ2025}
Nayana, A.~J., Margutti, R., Wiston, E., {et~al.} 2025,
  \bibinfo{title}{Dinosaur in a Haystack: X-Ray View of the Entrails of SN
  2023ixf and the Radio Afterglow of Its Interaction with the Medium Spawned by
  the Progenitor Star (Paper I)*,} The Astrophysical Journal, 985, 51,
  \dodoi{10.3847/1538-4357/adc2fb}

\bibitem[{T. pandas~development team(2020)pandas~development
  team}]{reback2020pandas}
pandas~development team, T. 2020, \bibinfo{title}{pandas-dev/pandas: Pandas,},
  2.2.3 Zenodo, \dodoi{10.5281/zenodo.3509134}

\bibitem[{S. {Panjkov} {et~al.}(2024){Panjkov}, {Auchettl}, {Shappee}, {Do},
  {Lopez}, \& {Beacom}}]{Panjkov24}
{Panjkov}, S., {Auchettl}, K., {Shappee}, B.~J., {et~al.} 2024,
  \bibinfo{title}{{Probing the soft X-ray properties and multi-wavelength
  variability of SN2023ixf and its progenitor},} \pasa, 41, e059,
  \dodoi{10.1017/pasa.2024.66}

\bibitem[{D. {Pooley} \& W.~H.~G. {Lewin}(2002){Pooley} \&
  {Lewin}}]{2002IAUC.8024....2P}
{Pooley}, D., \& {Lewin}, W.~H.~G. 2002, \bibinfo{title}{{Supernova 2002hh in
  NGC 6946},} \iaucirc, 8024, 2

\bibitem[{D. {Pooley} \& W.~H.~G. {Lewin}(2004){Pooley} \&
  {Lewin}}]{2004IAUC.8390....1P}
{Pooley}, D., \& {Lewin}, W.~H.~G. 2004, \bibinfo{title}{{Supernova 2004dj in
  NGC 2403},} \iaucirc, 8390, 1

\bibitem[{D. Pooley {et~al.}(2002)Pooley, Lewin, Fox, Miller, Lacey, Van~Dyk,
  Weiler, Sramek, Filippenko, Leonard, Immler, Chevalier, Fabian, Fransson, \&
  Nomoto}]{Pooley_2002}
Pooley, D., Lewin, W. H.~G., Fox, D.~W., {et~al.} 2002, \bibinfo{title}{X-Ray,
  Optical, and Radio Observations of the Type II Supernovae 1999em and 1998S,}
  The Astrophysical Journal, 572, 932, \dodoi{10.1086/340346}

\bibitem[{F.~D. {Romanov}(2024){Romanov}}]{2024TNSAN.109....1R}
{Romanov}, F.~D. 2024, \bibinfo{title}{{Follow-up observation of SN 2024ggi in
  NGC 3621},} Transient Name Server AstroNote, 109, 1

\bibitem[{P.~W.~A. {Roming} {et~al.}(2009){Roming}, {Pritchard}, {Brown},
  {Holland}, {Immler}, {Stockdale}, {Weiler}, {Panagia}, {Van Dyk},
  {Hoversten}, {Milne}, {Oates}, {Russell}, \&
  {Vandrevala}}]{2009ApJ...704L.118R}
{Roming}, P.~W.~A., {Pritchard}, T.~A., {Brown}, P.~J., {et~al.} 2009,
  \bibinfo{title}{{Multi-Wavelength Properties of the Type IIb SN 2008ax},}
  \apjl, 704, L118, \dodoi{10.1088/0004-637X/704/2/L118}

\bibitem[{S. {Ryder} {et~al.}(2024){Ryder}, {Maeda}, {Chandra}, {Alsaberi}, \&
  {Kotak}}]{2024ATel16616....1R}
{Ryder}, S., {Maeda}, K., {Chandra}, P., {Alsaberi}, R., \& {Kotak}, R. 2024,
  \bibinfo{title}{{Radio detection of SN 2024ggi},} The Astronomer's Telegram,
  16616, 1

\bibitem[{A. Saha {et~al.}(2006)Saha, Thim, Tammann, Reindl, \&
  Sandage}]{Saha_2006}
Saha, A., Thim, F., Tammann, G.~A., Reindl, B., \& Sandage, A. 2006,
  \bibinfo{title}{Cepheid Distances to SNe Ia Host Galaxies Based on a Revised
  Photometric Zero Point of the HST WFPC2 and New PL Relations and Metallicity
  Corrections,} The Astrophysical Journal Supplement Series, 165, 108,
  \dodoi{10.1086/503800}

\bibitem[{ {SAS development team}(2014){SAS development
  team}}]{2014ascl.soft04004S}
{SAS development team}. 2014, \bibinfo{title}{{SAS: Science Analysis System for
  XMM-Newton observatory},}, Astrophysics Source Code Library, record
  ascl:1404.004

\bibitem[{E.~M. Schlegel(1999)Schlegel}]{Schlegel_1999}
Schlegel, E.~M. 1999, \bibinfo{title}{X-Ray Detection of SN 1994W in NGC 4041?}
  The Astrophysical Journal, 527, L85, \dodoi{10.1086/312408}

\bibitem[{E.~M. Schlegel(2001)Schlegel}]{Schlegel_2001}
Schlegel, E.~M. 2001, \bibinfo{title}{Chandra Observations of SN 1999gi and the
  X-Ray Emission of Type II-P Supernovae,} The Astrophysical Journal, 556, L25,
  \dodoi{10.1086/322269}

\bibitem[{M. {Shrestha} {et~al.}(2024){Shrestha}, {Bostroem}, {Sand},
  {Hosseinzadeh}, {Andrews}, {Dong}, {Hoang}, {Janzen}, {Pearson}, {Jencson},
  {Lundquist}, {Mehta}, {Ravi}, {Meza Retamal}, {Valenti}, {Brown}, {Jha},
  {Macrie}, {Hsu}, {Farah}, {Howell}, {McCully}, {Newsome}, {Padilla Gonzalez},
  {Pellegrino}, {Terreran}, {Kwok}, {Smith}, {Schwab}, {Martas}, {Munoz},
  {Medina}, {Li}, {Diaz}, {Hiramatsu}, {Tucker}, {Wheeler}, {Wang}, {Zhai},
  {Zhang}, {Gangopadhyay}, {Yang}, \& {Guti{\'e}rrez}}]{2024ApJ...972L..15S}
{Shrestha}, M., {Bostroem}, K.~A., {Sand}, D.~J., {et~al.} 2024,
  \bibinfo{title}{{Extended Shock Breakout and Early Circumstellar Interaction
  in SN 2024ggi},} \apjl, 972, L15, \dodoi{10.3847/2041-8213/ad6907}

\bibitem[{A. {Singh} {et~al.}(2024){Singh}, {Teja}, {Moriya}, {Maeda},
  {Kawabata}, {Tanaka}, {Imazawa}, {Nakaoka}, {Gangopadhyay}, {Yamanaka},
  {Swain}, {Sahu}, {Anupama}, {Kumar}, {Anche}, {Sano}, {Raj}, {Agnihotri},
  {Bhalerao}, {Bisht}, {Bisht}, {Belwal}, {Chakrabarti}, {Fujii}, {Nagayama},
  {Matsumoto}, {Hamada}, {Kawabata}, {Kumar}, {Kumar}, {Malkan}, {Smith},
  {Sakagami}, {Taguchi}, {Tominaga}, \& {Watanabe}}]{2024ApJ...975..132S}
{Singh}, A., {Teja}, R.~S., {Moriya}, T.~J., {et~al.} 2024,
  \bibinfo{title}{{Unravelling the Asphericities in the Explosion and
  Multifaceted Circumstellar Matter of SN 2023ixf},} \apj, 975, 132,
  \dodoi{10.3847/1538-4357/ad7955}

\bibitem[{S. {Srivastav} {et~al.}(2024){Srivastav}, {Chen}, {Smartt},
  {Nicholl}, {Smith}, {Young}, {Fulton}, {McCollum}, {Moore}, {Weston},
  {Sheng}, {Aamer}, {Angus}, {Ramsden}, {Shingles}, {Gillanders}, {Rhodes},
  {Andersson}, {Stevance}, {Denneau}, {Tonry}, {Weiland}, {Lawrence}, {Siverd},
  {Erasmus}, {Koorts}, {Jordan}, {Suc}, {Rest}, {Stubbs}, \&
  {Sommer}}]{2024TNSAN.100....1S}
{Srivastav}, S., {Chen}, T.~W., {Smartt}, S.~J., {et~al.} 2024,
  \bibinfo{title}{{ATLAS24fsk (AT2024ggi): discovery of a nearby candidate SN
  in NGC 3621 at 7 Mpc with a possible progenitor detection},} Transient Name
  Server AstroNote, 100, 1

\bibitem[{T. Szalai {et~al.}(2019)Szalai, Vinkó, Könyves-Tóth, Nagy,
  Bostroem, Sárneczky, Brown, Pejcha, Bódi, Cseh, Csörnyei, Dencs, Hanyecz,
  Ignácz, Kalup, Kriskovics, Ordasi, Pál, Seli, Sódor, Szakáts, Vida,
  Zsidi, team, Arcavi, Ashall, Burke, Galbany, Hiramatsu, Hosseinzadeh, Hsiao,
  Howell, McCully, Moran, Rho, Sand, Shahbandeh, Valenti, Wang, Wheeler, \&
  Project}]{Szalai_2019}
Szalai, T., Vinkó, J., Könyves-Tóth, R., {et~al.} 2019, \bibinfo{title}{The
  Type II-P Supernova 2017eaw: From Explosion to the Nebular Phase,} The
  Astrophysical Journal, 876, 19, \dodoi{10.3847/1538-4357/ab12d0}

\bibitem[{B.~P. {Thomas} {et~al.}(2022){Thomas}, {Wheeler}, {Dwarkadas},
  {Stockdale}, {Vink{\'o}}, {Pooley}, {Xu}, {Zeimann}, \&
  {MacQueen}}]{2022ApJ...930...57T}
{Thomas}, B.~P., {Wheeler}, J.~C., {Dwarkadas}, V.~V., {et~al.} 2022,
  \bibinfo{title}{{Seven Years of SN 2014C: A Multiwavelength Synthesis of an
  Extraordinary Supernova},} \apj, 930, 57, \dodoi{10.3847/1538-4357/ac5fa6}

\bibitem[{J. {Tonry} {et~al.}(2024){Tonry}, {Denneau}, {Weiland}, {Lawrence},
  {Siverd}, {Erasmus}, {Koorts}, {Jordan}, {Suc}, {Smartt}, {Smith}, {Young},
  {Nicholl}, {Fulton}, {McCollum}, {Moore}, {Weston}, {Sheng}, {Ramsden},
  {Angus}, {Aamer}, {Shingles}, {Srivastav}, {Gillanders}, {Rhodes},
  {Andersson}, {Stevance}, {Rest}, {Chen}, {Stubbs}, \&
  {Sommer}}]{2024TNSTR1020....1T}
{Tonry}, J., {Denneau}, L., {Weiland}, H., {et~al.} 2024,
  \bibinfo{title}{{ATLAS Transient Discovery Report for 2024-04-11},} Transient
  Name Server Discovery Report, 2024-1020, 1

\bibitem[{D.~J. {Turner} {et~al.}(2022){Turner}, {Giles}, {Romer}, \&
  {Korbina}}]{2022arXiv220201236T}
{Turner}, D.~J., {Giles}, P.~A., {Romer}, A.~K., \& {Korbina}, V. 2022,
  \bibinfo{title}{{XGA: A module for the large-scale scientific exploitation of
  archival X-ray astronomy data},} arXiv e-prints, arXiv:2202.01236,
  \dodoi{10.48550/arXiv.2202.01236}

\bibitem[{S.~S. {Vasylyev} {et~al.}(2025){Vasylyev}, {Dessart}, {Yang},
  {Filippenko}, {Patra}, {Brink}, {Wang}, {Chornock}, {Margutti}, {Gates},
  {Burgasser}, {Sears}, {Karpoor}, {LeBaron}, {Softich}, {Theissen}, {Wiston},
  \& {Zheng}}]{2025arXiv250503975V}
{Vasylyev}, S.~S., {Dessart}, L., {Yang}, Y., {et~al.} 2025,
  \bibinfo{title}{{Spectropolarimetric Evolution of SN 2023ixf: an Asymmetric
  Explosion in a Confined Aspherical Circumstellar Medium},} arXiv e-prints,
  arXiv:2505.03975, \dodoi{10.48550/arXiv.2505.03975}

\bibitem[{J. {Vink}(2012){Vink}}]{2012A&ARv..20...49V}
{Vink}, J. 2012, \bibinfo{title}{{Supernova remnants: the X-ray perspective},}
  \aapr, 20, 49, \dodoi{10.1007/s00159-011-0049-1}

\bibitem[{P. Virtanen {et~al.}(2020)Virtanen, Gommers, Oliphant, Haberland,
  Reddy, Cournapeau, Burovski, Peterson, Weckesser, Bright, {van der Walt},
  Brett, Wilson, Millman, Mayorov, Nelson, Jones, Kern, Larson, Carey, Polat,
  Feng, Moore, {VanderPlas}, Laxalde, Perktold, Cimrman, Henriksen, Quintero,
  Harris, Archibald, Ribeiro, Pedregosa, {van Mulbregt}, \& {SciPy 1.0
  Contributors}}]{2020SciPy-NMeth}
Virtanen, P., Gommers, R., Oliphant, T.~E., {et~al.} 2020,
  \bibinfo{title}{{{SciPy} 1.0: Fundamental Algorithms for Scientific Computing
  in Python},} Nature Methods, 17, 261, \dodoi{10.1038/s41592-019-0686-2}

\bibitem[{ {W}es {M}c{K}inney(2010){W}es
  {M}c{K}inney}]{mckinney-proc-scipy-2010}
{W}es {M}c{K}inney. 2010, in {P}roceedings of the 9th {P}ython in {S}cience
  {C}onference, ed. {S}t\'efan van~der {W}alt \& {J}arrod {M}illman, 56 -- 61,
  \dodoi{10.25080/Majora-92bf1922-00a}

\bibitem[{L. {Wyrzykowski} {et~al.}(2025){Wyrzykowski}, {Mikolajczyk},
  {Kotysz}, {Zielinski}, {Hambsch}, \& {Bronikowski}}]{2025TNSAN..22....1W}
{Wyrzykowski}, L., {Mikolajczyk}, P., {Kotysz}, K., {et~al.} 2025,
  \bibinfo{title}{{Photometric follow-up of SN2024ggi with BHTOM.space global
  telescope network},} Transient Name Server AstroNote, 22, 1

\bibitem[{D. {Xiang} {et~al.}(2024){Xiang}, {Mo}, {Wang}, {Wang}, {Zhang},
  {Lin}, {Chen}, {Song}, {Liu}, {Wang}, \& {Li}}]{2024ApJ...969L..15X}
{Xiang}, D., {Mo}, J., {Wang}, X., {et~al.} 2024, \bibinfo{title}{{The Red
  Supergiant Progenitor of Type II Supernova 2024ggi},} \apjl, 969, L15,
  \dodoi{10.3847/2041-8213/ad54b3}

\bibitem[{Q. {Zhai} {et~al.}(2024){Zhai}, {Li}, {Zhang}, \&
  {Wang}}]{2024TNSCR1031....1Z}
{Zhai}, Q., {Li}, L., {Zhang}, J., \& {Wang}, X. 2024, \bibinfo{title}{{LiONS
  Transient Classification Report for 2024-04-11},} Transient Name Server
  Classification Report, 2024-1031, 1

\bibitem[{J. {Zhang} {et~al.}(2024{\natexlab{a}}){Zhang}, {Li}, {Cheng}, {Wu},
  {Jia}, {Chen}, {Cui}, {Feng}, {Guan}, {Han}, {Li}, {Liu}, {Lu}, {Song},
  {Wang}, {Xu}, {Zhang}, {Zhao}, {Zhao}, {Jin}, {Ling}, {Liu}, {Liu}, {Liu},
  {Li}, {Sun}, {Yuan}, {Zhang}, {Zhang}, {Li}, {Wang}, {Zhou}, {Nandra}, {Rau},
  {Friedrich}, {Meidinger}, {Burwitz}, {Kuulkers}, {Santovincenzo}, {O'Brien},
  {Cordier}, {Wang}, \& {Li}}]{2024ATel16588....1Z}
{Zhang}, J., {Li}, C.~K., {Cheng}, H.~Q., {et~al.} 2024{\natexlab{a}},
  \bibinfo{title}{{SN 2024ggi: detection of X-ray emission by EP-FXT},} The
  Astronomer's Telegram, 16588, 1

\bibitem[{J. {Zhang} {et~al.}(2024{\natexlab{b}}){Zhang}, {Dessart}, {Wang},
  {Zhai}, {Yang}, {Li}, {Lin}, {Valerin}, {Cai}, {Guo}, {Wang}, {Zhao}, {Wang},
  \& {Yan}}]{2024ApJ...970L..18Z}
{Zhang}, J., {Dessart}, L., {Wang}, X., {et~al.} 2024{\natexlab{b}},
  \bibinfo{title}{{Probing the Shock Breakout Signal of SN 2024ggi from the
  Transformation of Early Flash Spectroscopy},} \apjl, 970, L18,
  \dodoi{10.3847/2041-8213/ad5da4}

\bibitem[{E.~A. {Zimmerman} {et~al.}(2024){Zimmerman}, {Irani}, {Chen},
  {Gal-Yam}, {Schulze}, {Perley}, {Sollerman}, {Filippenko}, {Shenar}, {Yaron},
  {Shahaf}, {Bruch}, {Ofek}, {De Cia}, {Brink}, {Yang}, {Vasylyev}, {Ben Ami},
  {Aubert}, {Badash}, {Bloom}, {Brown}, {De}, {Dimitriadis}, {Fransson},
  {Fremling}, {Hinds}, {Horesh}, {Johansson}, {Kasliwal}, {Kulkarni},
  {Kushnir}, {Martin}, {Matuzewski}, {McGurk}, {Miller}, {Morag}, {Neil},
  {Nugent}, {Post}, {Prusinski}, {Qin}, {Raichoor}, {Riddle}, {Rowe},
  {Rusholme}, {Sfaradi}, {Sjoberg}, {Soumagnac}, {Stein}, {Strotjohann},
  {Terwel}, {Wasserman}, {Wise}, {Wold}, {Yan}, \&
  {Zhang}}]{2024Natur.627..759Z}
{Zimmerman}, E.~A., {Irani}, I., {Chen}, P., {et~al.} 2024,
  \bibinfo{title}{{The complex circumstellar environment of supernova
  2023ixf},} \nat, 627, 759, \dodoi{10.1038/s41586-024-07116-6}

\end{thebibliography}
\bibliographystyle{aasjournalv7}

\appendix 
\vspace{-0.25cm}
\restartappendixnumbering

\section{Contour Plot of Best Fit Parameter}

\begin{figure*}[htb!]
  \gridline{
    \fig{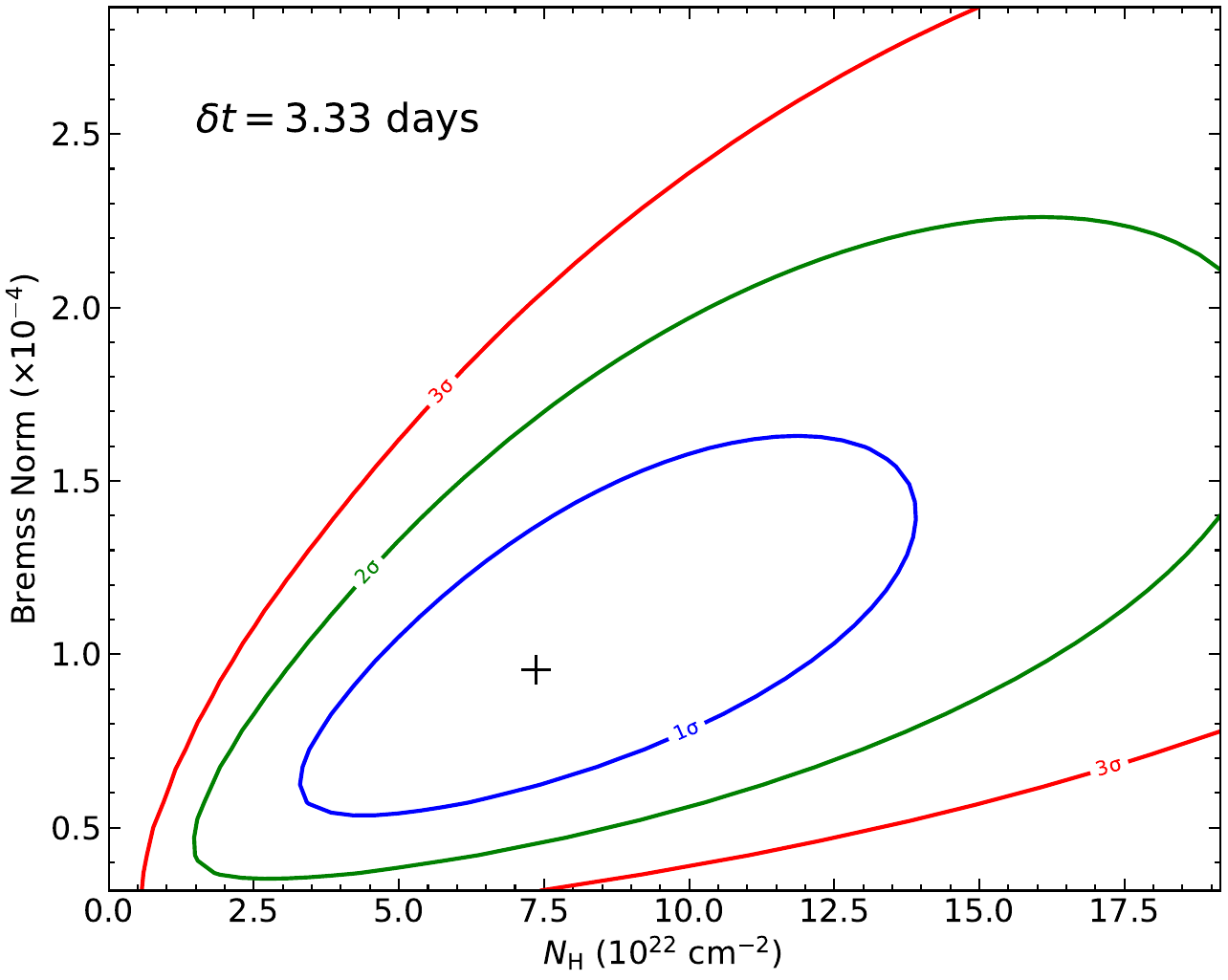}{0.5\textwidth}{}
    \fig{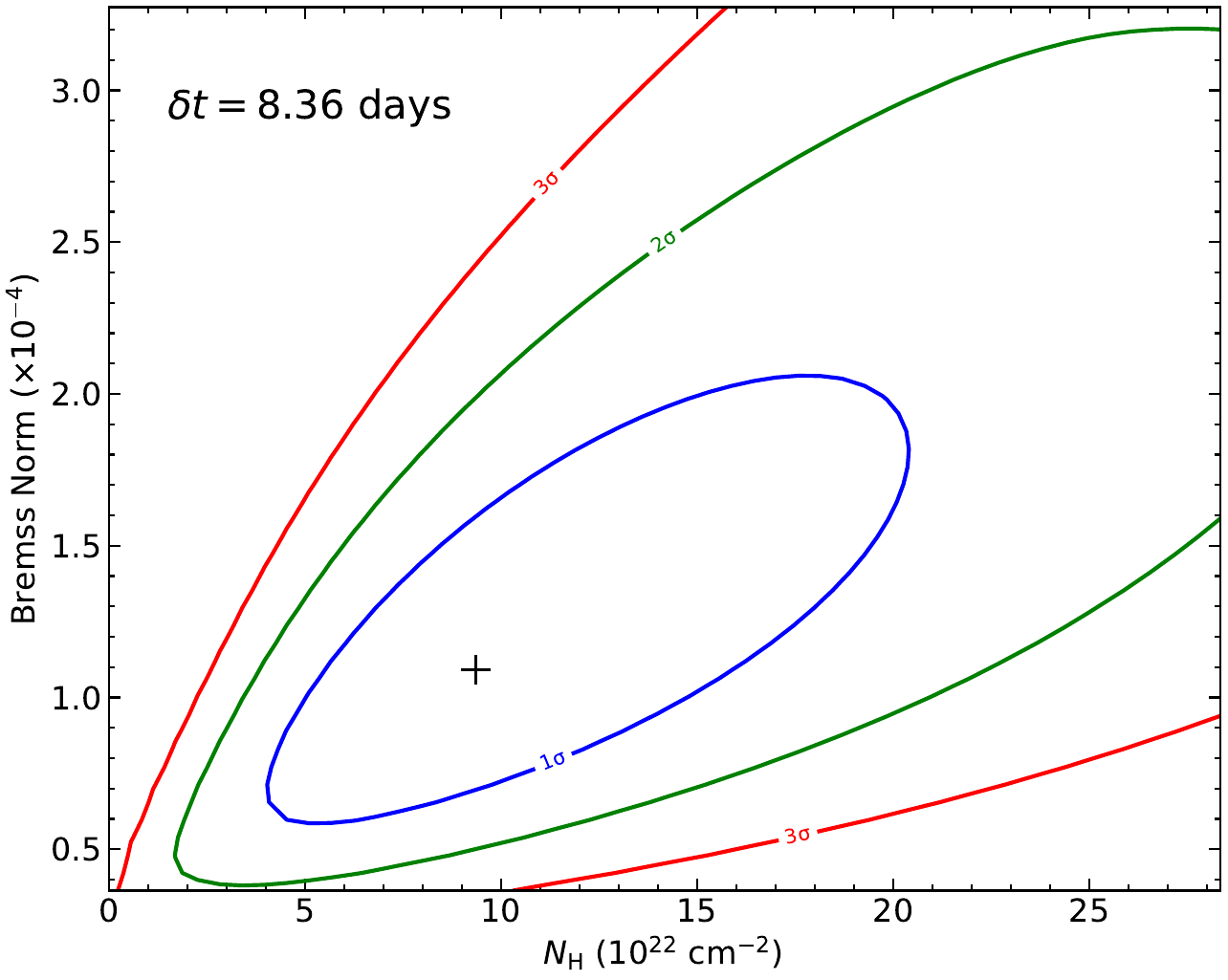}{0.5\textwidth}{}
  }
  \gridline{
    \fig{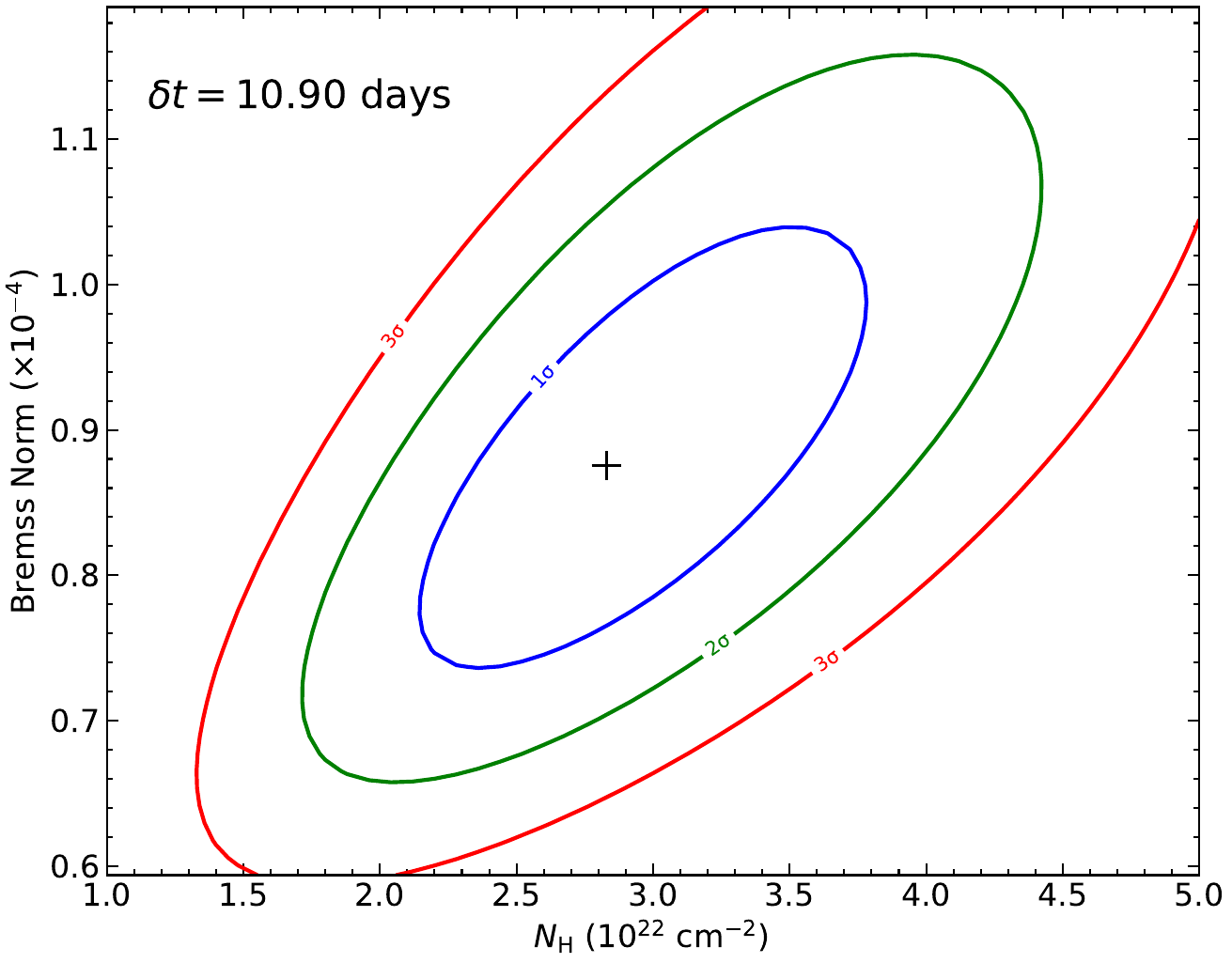}{0.5\textwidth}{}
    \fig{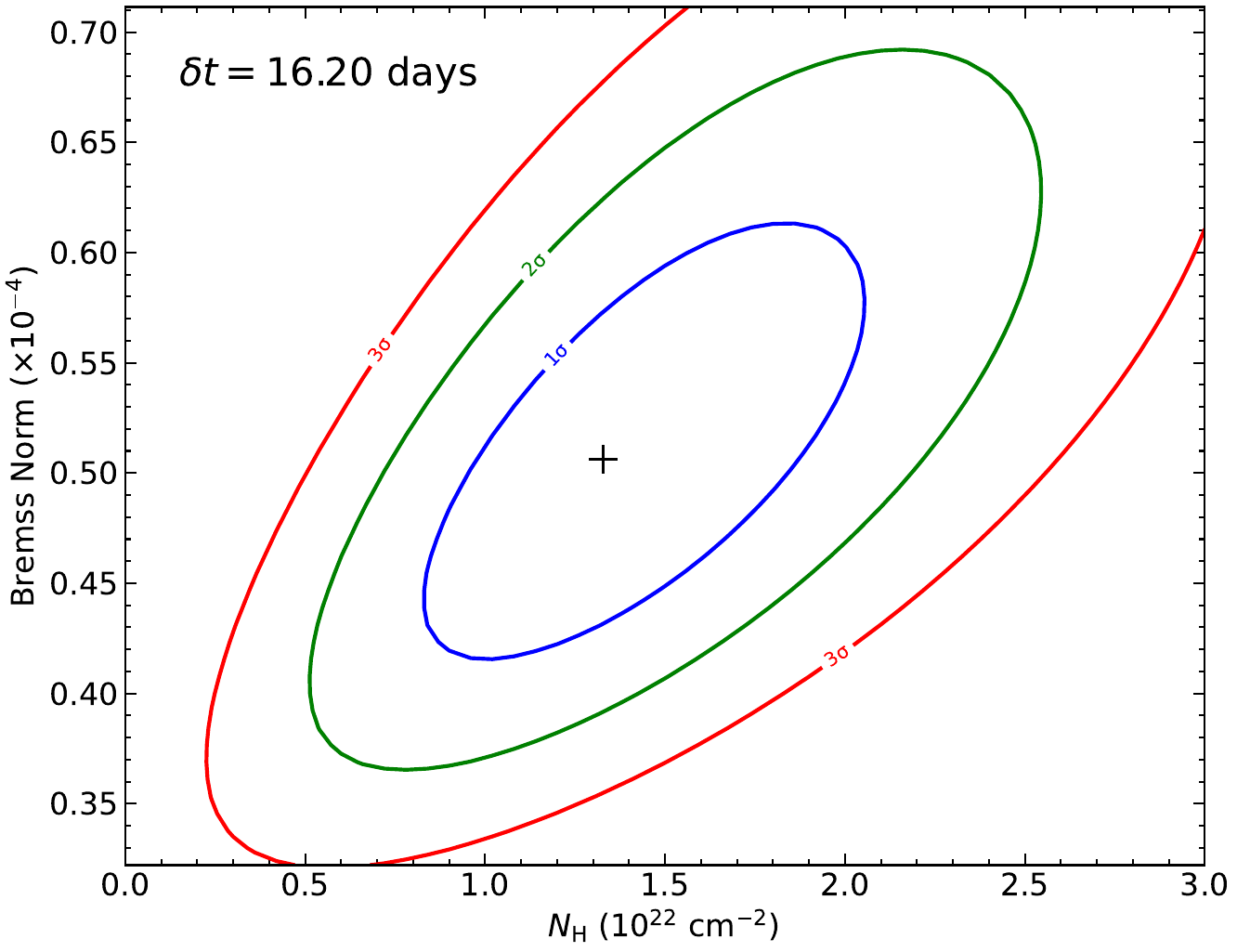}{0.5\textwidth}{}
  }
  
  \caption{\textit{Top left:} Contour plot of the best fit parameter values for the column density and the Bremsstrahlung normalization from XRT data at $\delta t = 3.33 \ \rm days$. \textit{Top right:} Contour plot of the best fit parameter values for the column density and the Bremsstrahlung normalization from XRT data at $\delta t = 8.36 \ \rm days$. \textit{Bottom left:} Contour plot of the best fit parameter values for the column density and the Bremsstrahlung normalization from CXO data at $\delta t = 10.90 \ \rm days$. \textit{Bottom right:} Contour plot of the best fit parameter values for the column density and the Bremsstrahlung normalization from CXO data at $\delta t = 16.20 \ \rm days$. Both parameters are well constrained in all the epochs, i.e. at $\delta t = 3.33 \ \rm days$, $\delta t = 8.36 \ \rm days$, $\delta t = 10.90 \ \rm days$, $\delta t = 16.20 \ \rm days$. The blue line represents $1\sigma$ range, the green line represents $2\sigma$ range, and the red line represents $3\sigma$ range.}
  \label{fig:contour}
\end{figure*}

\section{X-Ray Observation Logs}

\startlongtable
\begin{deluxetable}{c c c c c c}
	\tablecaption{X-ray Observation Log of SN~2024ggi\label{tab:obslogggi}}
	\tabletypesize{\small}
	\setlength{\tabcolsep}{6pt}
	\tablehead{
		\colhead{\textbf{Instrument}} &
		\colhead{\textbf{Observation}} &
		\colhead{\textbf{Mid}} &
		\colhead{\textbf{Observation ID}} &
		\colhead{\textbf{Exposure}} &
		\colhead{\textbf{PI}} \\[-6pt]
		\colhead{} &
		\colhead{\textbf{Date}} &
		\colhead{\textbf{Time}\tablenotemark{a}} &
		\colhead{} &
		\colhead{\textbf{Time} (ks)} &
		\colhead{}\\[-6pt]
		\colhead{} &
		\colhead{(yyyy/mm/dd)} &
		\colhead{(days)} &
		\colhead{} &
		\colhead{} &
		\colhead{}
	}
	\startdata
	Swift-XRT &	2024-04-11 -- &	3.33	& 00045607063, 00016601001, 00016601002,  &	20	& Swift, \\[-3pt]
	& 2024-04-16 & & 00016601003, 00016601004, 00016601005, & & D. Sand \\[-3pt]
	& & & 00016601006, 00016601007, 00016601012, & & \\[-3pt]
	& & & 00016601013, 00016601014, 00016601015, & & \\[-3pt]
	& & & 00016601017, 00016601018, 00016601019, & & \\[-3pt]
	& & & 00045607064, 00045607065 & & \\
	Swift-XRT &	2024-04-17 --	& 8.36	& 00016601020, 00016601022, 00016601023, &	9	& Swift, \\[-3pt]
	& 2024-04-20 & & 00016601024, 00016601025, 00016601026, & & D. Sand \\[-3pt]
	& & & 00016601027 & & \\
	CXO         & 2024-04-21                & 10.9  & 29383                                                                     & 14.9 & E.\ Zimmerman \\
	CXO         & 2024-04-26                & 16.2  & 29384                                                                     & 14.6 & E.\ Zimmerman \\
	Swift‐XRT   & 2024-05-01 -- & 24.49 &
	00016601044, 00016601045, 00016601048, &
	7.6  & D. Sand \\[-3pt]
	& 2024-05-09 & & 00016601049, 00016601050,
	00016601051, & & \\[-3pt]
	& & & 00016601052, 00016601053 & & \\
	Swift‐XRT   & 2024-05-12 -- & 34.76 &
	00016601055, 00016601059, 00016601062, &
	5.0  & D. Sand \\[-3pt]
	& 2024-05-18 & & 00016601063 & & \\
	Swift‐XRT   & 2024-05-21 -- & 43.2  &
	00016601064, 00016601065, 00016601066, &
	10.1 & D. Sand \\[-3pt]
	& 2024-05-26 & & 00016601067, 00016601068 & & \\
	XMM/EPIC-pn & 2024-06-04                & 55.03 & 0882480901                                                               & 16.9 & W.\ Jacobson-Galan \\
	XMM/EPIC-MOS1 & 2024-06-04              & 55.03 & 0882480901                                                               & 16.2 & W.\ Jacobson-Galan \\
	XMM/EPIC-MOS2 & 2024-06-04              & 55.03 & 0882480901                                                               & 16.2 & W.\ Jacobson-Galan \\
	Swift‐XRT   & 2024-06-02 -- & 55.7  &
	00016601069, 00016601070, 00016601071 &
	2.4  & D. Sand \\[-3pt]
	& 2024-06-08 & & & & \\
	Swift‐XRT   & 2024-06-24 -- & 81.51 &
	00016601072, 00016601073, 00016601074 &
	6.1  & D. Sand \\[-3pt]
	& 2024-07-08 & & & & \\
	XMM/EPIC-pn & 2024-07-05                & 85.4  & 0882481001                                                               & 9.8  & W.\ Jacobson-Galan \\
	XMM/EPIC-MOS1 & 2024-07-05              & 85.4  & 0882481001                                                               & 11.6 & W.\ Jacobson-Galan \\
	XMM/EPIC-MOS2 & 2024-07-05              & 85.4  & 0882481001                                                               & 11.6 & W.\ Jacobson-Galan \\
	Swift‐XRT   & 2024-07-15 -- & 106.35 &
	00016601075, 00016601076, 00016601077, &
	4.7  & D. Sand \\[-3pt]
	& 2024-08-05 & & 00016601078  & & \\
	Swift‐XRT   & 2024-08-08 -- & 125.21 &
	00016601079, 00016601080, 00016601081 &
	4.9  & D. Sand \\[-3pt]
	& 2024-08-19 & & & & \\
	Swift‐XRT   & 2024-11-04                & 208.13 & 00045607066                                                              & 2.8  & Swift \\
	Swift‐XRT   & 2024-12-21 -- & 260.33 &
	00045607067, 00045607068, 00045607069, &
	5.2  & Swift \\[-3pt]
	& 2025-01-01 & & 00045607070 & & \\
	Swift‐XRT   & 2025-01-08 -- & 276.65 &
	00045607071, 00045607072, 00045607075 &
	4.0  & Swift \\[-3pt]
	& 2025-01-15 & & & & \\
	Swift‐XRT   & 2025-02-08 -- & 316.43 &
	00045607076, 00045607078, 00045607080 &
	3.7  & Swift \\[-3pt]
	& 2025-03-05 & & & & \\
	Swift‐XRT   & 2025-03-10 -- & 338.76 &
	00045607081, 00045607082 &
	2.9  & Swift \\[-3pt]
	& 2025-03-20 & & & & \\[3pt]
	\enddata
	\tablenotetext{a}{With respect to time of explosion.}
\end{deluxetable}

\end{document}